\documentclass{article}

\usepackage{PRIMEarxiv}

\usepackage[utf8]{inputenc} 
\usepackage[T1]{fontenc}    
\usepackage{hyperref}       
\usepackage{url}            
\usepackage{booktabs}       
\usepackage{amsfonts}       
\usepackage{nicefrac}       
\usepackage{microtype}      
\usepackage{lipsum}
\usepackage{fancyhdr}       
\usepackage{graphicx}       
\graphicspath{{media/}}     
\usepackage{amsmath}
\usepackage{amsfonts}
\usepackage{amssymb}
\usepackage{geometry}
\usepackage{graphics}
\usepackage{bm}
\usepackage{booktabs}
\usepackage{multirow}
\usepackage{caption}
\usepackage{subcaption}

\usepackage[flushleft]{threeparttable}

\usepackage{algorithm}
\usepackage{algpseudocode}

\title{MAntRA: A framework for model agnostic reliability analysis}

\author{
  Yogesh Chandrakant Mathpati \\
  Department of Applied Mechanics \\
  Indian Institute of Technology Delhi \\
  \texttt{ama212637@iitd.ac.in} \\
   \And
  Kalpesh Sanjay More \\
  Department of Applied Mechanics \\
  Indian Institute of Technology Delhi \\
  \texttt{ama212077@iitd.ac.in}\\
  \AND
  Tapas Tripura \\
  Department of Applied Mechanics \\
 Indian Institute of Technology Delhi
\\
  \texttt{tapas.t@am.iitd.ac.in} \\
  \And
  Rajdip Nayek \\
  Department of Applied Mechanics \\
 Indian Institute of Technology Delhi
\\
  \texttt{rajdipn@am.iitd.ac.in} \\
  \And
  Souvik Chakraborty \\
  Department of Applied Mechanics \\
  Yardi School of Artificial Intelligence (ScAI)
 \\
 Indian Institute of Technology Delhi
\\
  \texttt{souvik@am.iitd.ac.in} \\
}

\begin{document}
\maketitle

\begin{abstract}
We propose a novel model agnostic data-driven reliability analysis framework for time-dependent reliability analysis. The proposed approach -- referred to as MAntRA -- combines interpretable machine learning, Bayesian statistics, and identifying stochastic dynamic equation to evaluate reliability of stochastically-excited dynamical systems for which the governing physics is \textit{apriori} unknown. A two-stage approach is adopted: in the first stage, an efficient variational Bayesian equation discovery algorithm is developed to determine the governing physics of an underlying stochastic differential equation (SDE) from measured output data. The developed algorithm is efficient and accounts for epistemic uncertainty due to limited and noisy data, and aleatoric uncertainty because of environmental effect and external excitation. In the second stage, the discovered SDE is solved using a stochastic integration scheme and the probability failure is computed. The efficacy of the proposed approach is illustrated on three numerical examples. The results obtained indicate the possible application of the proposed approach for reliability analysis of in-situ and heritage structures from on-site measurements. 
\end{abstract}












\section{Introduction}
Reliability analysis of uncertain dynamical systems stands as one of the most important problems in engineering, as it qualifies the safety of an engineering system subject to variability in its input conditions. Conventionally, reliability analysis is carried out at the design stage by exploiting the known model of the system and statistical computing. In literature, there exists a plethora of methods including first-order reliability method (FORM) \cite{hohenbichler1987new} and second-order reliability method (SORM) \cite{der1987second,kiureghian1991efficient,adhikari2004reliability}, Monte Carlo simulation (MCS) \cite{thakur1978monte,boyaval2012fast,tamimi1989monte}, importance sampling \cite{au1999new} and subset simulation \cite{au2001estimation} for solving reliability analysis at the design stage. However, an equally important problem that is often neglected is reliability analysis of existing systems, particularly the heritage structures. It is well acknowledged that structures undergo degradation which can alter the governing physics of the system. Under such circumstances, it is not possible estimate reliability of a system using the traditional methods where a model based on design blueprint of the structure is used. In this regard, the objective of this paper is to address this apparent limitation by developing a model-agnostic framework that can be used for evaluating reliability of systems with unknown physics. 

One potential direction for developing model-agnostic reliability analysis framework is to employ surrogate modelling techniques. Surrogate modelling is already popular in the reliability analysis literature. The fundamental idea is to train a machine learning model by generating data from the known model and use it as a surrogate to the known model; this accelerates the overall process and allows solving reliability analysis problem in a reasonable time. Popular surrogate models existing in the literature includes response surface method \cite{zhang2017time,soares2002reliability} Gaussian process \cite{bilionis2012multi,peng2014inverse}, analysis-of-variance decomposition \cite{yang2012adaptive,chakraborty2017towards,chakraborty2016modelling}, polynomial chaos expansion{\cite{sudret2008global,xiu2002wiener}}, support vector machine {\cite{roy2019support,cheng2021adaptive}} and neural network \cite{elhewy2006reliability,hurtado2001neural} to name a few. In the context of systems with unknown governing physics, one possibility is to train a surrogate model using on-field measurement (or historical data if available). Unfortunately, the black-box surrogate models often used in reliability analysis do not generalize beyond the training window and hence, has limited applicability for systems with unknown physics.

Advancements in machine learning tools and new sophisticated data measurement devices have given rise to data-driven equation discovery methods. One of the most popular methods for equation discovery is Sparse identification of non-linear Dynamics (SINDy) \cite{brunton2016discovering}. The idea behind SINDy is to construct a library of candidate functions and then use sparse linear regression to select only relevant candidate functions to form the best-fit model. SINDy finds applications in many areas,  some of which includes sparse identification of biological networks in biology \cite{mangan2016inferring}, sparse identification of chemical reaction in chemistry \cite{hoffmann2019reactive},  sparse model selection of dynamical system using information criteria \cite{mangan2017model}, sparse identification for predictive control \cite{kaiser2018sparse},  identification of structured dynamical systems with limited data \cite{schaeffer2020extracting}, and discovery of partial differential equations \cite{rudy2017data,zhang2018robust}.
However, these equations discovery techniques fail for account for the uncertainties associated with noisy and limited data, and aleatoric uncertainties due to environmental effect and/or external load are not accounted for. The Bayesian equation discovery techniques proposed in \cite{nayek2021spike,fuentes2021equation} address the challenge associated with limited and noisy data. While \cite{fuentes2021equation} proposed a relevance vector machine based Bayesian approach for equation discovery, \cite{nayek2021spike} exploits spike and slab prior \cite{ishwaran2005spike} and Gibbs sampling \cite{george1993variable} for learning the equation while accounting for uncertainty due to noisy and sparse data. However, the effect of aleatoric uncertainty due to environmental effect and/or external load are not accounted for in these approaches as well.

In this work, we propose a model-agnostic reliability analysis framework that involving (a) discovering the governing physics from data and (b) application of conventional reliability analysis techniques on the discovered model for computing the probability of failure. However, the equation discovery techniques discovered before are not suitable for reliability analysis as the effect of aleatoric uncertainty due to environmental effect and/or external load are not accounted for; in other words, a framework capable of discovering \textit{stochastic differential equation} (SDE) from data is required \cite{boninsegna2018sparse,tripura2022model,tripura2023sparse}. To that end, we propose, for the first time, a variational approach for discovering SDE from data. This is an improvement over our recently published work \cite{tripura2023sparse} where It{\^o} calculus, Gibbs sampling and spike and slab prior were employed for identifying governing SDE. With the proposed variational Bayes, the SDE discovery is significantly accelerated. Once the underlying SDE is discovered using the proposed approach, we employ the stochastic integration scheme rooted in It{\^o} calculus for computing the probability of failure. The overall framework is referred to as the Model-Agnostic Reliability Analysis (MAntRA) framework. To the best of our knowledge, this is the first attempt towards developing a reliability analysis framework that is model-agnostic and can exploit field data directly for reliability assessment. MAntRA has the following distinctive characteristics:
\begin{itemize}
    \item \textbf{Computational Efficiency:} One key component of MAntRA is the variational approach for discovering SDE from data. Compared to the recently proposed Gibbs sampling based approach for discovering SDE \cite{tripura2023sparse}. the proposed approach is computationally efficient.
    \item \textbf{Predictive uncertainty:} The proposed approach is rooted in Bayesian statistics and hence, can quantify the uncertainty associated with limited and noisy data. This is particularly important as old and heritage structures are susceptible to failure and knowledge on predictive uncertainty can help in taking an informed decision. 
\end{itemize}

The rest of the paper is organized as follows. In Section \ref{prob state}, the problem statement has been formally defined. The proposed MAntRA is discussed in details in Section \ref{sec:bayesian}. Efficacy of the proposed approach has been illustrated with three numerical examples in Section \ref{problems}. Finally, Section \ref{conclusion} provides the concluding remarks.

\section{Problem statement}\label{prob state}
Consider a generalized stochastic dynamical system of the form,
\begin{equation}\label{eq:sde}
    \dot{\boldsymbol{X}}(t)=\bm{f}\left(\boldsymbol{X}(t), t\right)+\bm{g}\left(\boldsymbol{X}(t), t\right) \bm{\zeta}(t)
\end{equation}
where $\boldsymbol{X}(t)$ denotes the $\mathbb{R}^{m}$-dimensional states of the process, $\bm{f}\left(\boldsymbol{X}(t), t\right)$ is the $\mathbb{R}^{m}$-dimensional vector representing the deterministic dynamics of the underlying phenomenon, $\bm{g}\left(\boldsymbol{X}(t), t\right)$ is the $\mathbb{R}^{m \times n}$-dimensional diffusion matrix influencing the input, and $\bm{\zeta}(t)$ is the $\mathbb{R}^{n}$-dimensional stochastic input modeled as $n$-dimensional Gaussian white noise. The probability of failure of a system defined using Eq. \eqref{eq:sde} at a given time $\tau$ can be computed by first employing a stochastic integration scheme and then performing simple post-processing of the results by using the limit-state function, $\mathcal J \left( \zeta \right) = 0$. Popular numerical integration scheme available in the literature includes Euler Maruyama, Milstein, and It\^{o} Taylor's 3.0 schemes \cite{kloeden1992higher,oksendal2013stochastic}. We note that the reliability analysis literature is generally based on the assumption that the governing physics in Eq. \eqref{eq:sde} is known. However, for many systems in science and engineering, the governing physics is either not known or only known in an approximate manner; this is particularly true when dealing with heritage buildings. Under such circumstances, the existing methods in the literature cannot be directly used for reliability analysis.

To formally define the problem statement, let us consider $\mathcal D = \left\{\bm {\tilde X} ^{(j)}_{1:\tau}\right\}_{j=1}^{N_t}$ to be the noisy measurements of $\bm X$ in the time-interval $[0,\tau]$. Given the fact that the system is stochastic, we also assume that measurements for $N_t$ such trajectories are available. With this setup, the objective is to solve a time-dependent reliability analysis problem and compute the probability of failure, $P_f$ of the system in $[0,T]$ with $T>> \tau$. We note that the primary challenge stems from the fact that the governing physics is unknown, and hence, it is not possible to use conventional stochastic integration schemes for reliability analysis.

\section{MAntRA: The proposed model-agnostic reliability framework}\label{sec:bayesian}
In this section, we present the framework of MAntRA for solving the problem mentioned in Section \ref{prob state}. MAntRA has two steps; in the first step, the proposed approach exploits Bayesian statistics and a sparse learning algorithm to learn the underlying governing physics from the time-history measurements $\mathcal D$. The second step involves performing the time-dependent reliability analysis of the systems governed by the discovered physics.

\subsection{Equation discovery}
The non-linear stochastic dynamical system represented by the Eq. \eqref{eq:sde} can be expressed in terms of the first-order It\^{o} SDE as follows:
\begin{equation} \label{eq:goveq}
    d \boldsymbol{X}(t)=\boldsymbol{f}\left(\boldsymbol{X}(t), t\right) d t+\boldsymbol{g}\left(\boldsymbol{X}(t), t\right) d \boldsymbol{B}(t) ; \quad \boldsymbol{X}\left(t=t_{0}\right)=\boldsymbol{X}_{0} ; \quad t \in[0, T]
\end{equation}
Here $\boldsymbol{B}(t)$ is the $\mathbb{R}^{n}$-dimensional Brownian motion, whose generalized derivative is the white noise. It has the properties $\mathbb{E}[{B}(t)]=0$, and $\mathbb{E}[B(s), B(t)]=\min (s, t)$. Note $\mathbb{E}[\cdot]$ represents the expectation operator. Our aim is to discover the Eq. \eqref{eq:goveq} from state measurements $\bm{X}$ alone. It is straightforward to understand that to discover the Eq. \eqref{eq:goveq}, we need to identify the functions $\boldsymbol{f}\left(\boldsymbol{X}(t), t\right)$, and $\boldsymbol{g}\left(\boldsymbol{X}(t), t\right)$. One problem is that the functions are coupled, and we need to decouple them so that we can solve for $\boldsymbol{f}\left(\boldsymbol{X}(t), t\right)$, and $\boldsymbol{g}\left(\boldsymbol{X}(t), t\right)$ independently and simultaneously. In this context, the Kramers-Moyal expansion (a generalization of the Fokker-Planck equation, see \cite{risken1996fokker}) provides us a way to directly express the functions $\boldsymbol{f}\left(\cdot, \cdot \right)$, and $\boldsymbol{g}\left(\cdot, \cdot\right)$ in terms of the states $\bm{X}$. 

To derive the expressions, we consider the transition probability density $p(X, t)$=${\rm{P}}(X,t|X_0,t_0)$ of the solution of Eq. \eqref{eq:goveq}. For the transition density, the Kramers-Moyal expansion is written as \cite{risken1996fokker},
\begin{equation}\label{eq:kramers}
    \dfrac{{\partial P(X,t)}}{{\partial t}} = {\sum\limits_{l = 1}^\infty  {\left( { - \dfrac{\partial }{{\partial X}}} \right)} ^l}{D^{(l)}}(X,t)p(X,t).
\end{equation}
The coefficients ${D^{(l)}}(\cdot, \cdot)$ of the expansion are given as,
\begin{equation}\label{eq:kramer_moments}
    {D^{(l)}}(X) = {\dfrac{1}{{l !}}\mathop {\lim }\limits_{{\Delta t}  \to 0} \dfrac{1}{{\Delta t} }\left\langle {{{\left| {X(t + {\Delta t} ) - X(t)} \right|}^{l}}} \right\rangle }.
\end{equation}
Although the states are stochastic processes, the function $\boldsymbol{f}\left(\cdot, \cdot \right)$ is deterministic and has a finite variation (no quadratic variation). On the contrary, the Brownian motion $\boldsymbol{B}(t)$ has a non-vanishing quadratic variation (zero finite variation) \cite{hassler2016stochastic}. Thus, the function $\boldsymbol{f}\left(\cdot, \cdot \right)$ can be expressed as a finite variation of the states, and the function $\boldsymbol{g}\left(\cdot, \cdot \right)$ can be expressed in terms of the quadratic variation of the state measurements. Therefore, by putting $l=1$ for finite variation and $l=2$ for quadratic variation in the Eq. \eqref{eq:kramer_moments}, we obtain the following representations for the drift, and diffusion terms, respectively,
\begin{align}\label{eq:kramer1}
    &{\bm{f}_{i}}(\bm{X}(t), t)= \lim _{\Delta t \rightarrow 0} \frac{1}{\Delta t} \mathbb{E}\left[\bm{X}_{i}(t+\Delta t)-\bm{X}_{i}(t)\right]; \quad i=1,\ldots,m\\
    &{\mathbf{R}}_{i j}(\bm{X}(t), t)=\frac{1}{2} \lim _{\Delta t \rightarrow 0} \frac{1}{\Delta t} \mathbb{E}\left[(\bm{X}_{i}(t+\Delta t)-\bm{X}_{i}(t))(\bm{X}_{j}(t+\Delta t)-\bm{X}_{j}(t))\right]; \quad i,j=1,\ldots,m \label{eq:kramer2}
\end{align}
where ${\mathbf{R}} \in \mathbb{R}^{n \times n}$ is the diffusion covariance matrix ${\mathbf{R}} :=$ $\bm{g}\left(t, \boldsymbol{X}(t)\right) \bm{g}\left(t, \boldsymbol{X}(t)\right)^{T}$. We assume that the drift and the diffusion components can be expressed analytically using some basis functions. The basis functions can be of the forms such as polynomial, trigonometric, etc. An example of the basis functions is provided later in Eq. \eqref{eq:library}. Next, let $\ell_{k}\left(\boldsymbol{X}(t)\right), k=1, \ldots, K$ be a collection of all possible basis functions. 
We then represent the functions $\boldsymbol{f}\left(\cdot, \cdot \right)$ and $\boldsymbol{g}\left(\cdot, \cdot \right)$ as the weighted linear superposition of the basis functions as,

\begin{equation} \label{eq:lib}
    \begin{aligned}
        &{f}_{i}\left(\boldsymbol{X}(t)\right)\approx\theta_{i, 1}^{f} \ell_{1}^{f}\left(\boldsymbol{X}(t)\right)+\ldots+\theta_{i, k}^{f} \ell_{k}^{f}\left(\boldsymbol{X}(t)\right)+\ldots+\theta_{i, K}^{f} \ell_{K}^{f}\left(\boldsymbol{X}(t)\right) \\
        &{\Gamma}_{i j}\left(\boldsymbol{X}(t)\right)\approx\theta_{i j, 1}^{g} \ell_{1}^{g}\left(\boldsymbol{X}(t)\right)+\ldots+\theta_{i j, k}^{g} \ell_{k}^{g}\left(\boldsymbol{X}(t)\right)+\ldots+\theta_{i j, K}^{g} \ell_{K}^{g}\left(\boldsymbol{X}(t)\right)
    \end{aligned}
\end{equation}
where  $\ell_{k}^{f}$ and $\ell_{k}^{g}$ are the basis functions for $\boldsymbol{f}\left(\cdot, \cdot \right)$ and $\boldsymbol{g}\left(\cdot, \cdot \right)$, respectively, and $\theta_{i, k}^{f}$ and $\theta_{i j, k}^{g}$ are the associated weights. For matrix representation, we can further define the dictionary functions $\mathbf{L}^{f} \in \mathbb{R}^{N \times K}$ and $\mathbf{L}^{g} \in \mathbb{R}^{N \times K}$ for $\boldsymbol{f}\left(\cdot, \cdot \right)$ and $\boldsymbol{g}\left(\cdot, \cdot \right)$, respectively, which contains all the candidate basis functions. Similarly, we define the weight vectors $\bm\theta^f$ for $\boldsymbol{f}\left(\cdot, \cdot \right)$ and $\bm\theta^g$ for $\boldsymbol{g}\left(\cdot, \cdot \right)$ as follows,
\begin{equation}
    \begin{aligned}
        \bm\theta^f &= [\theta_{i, 1}^{f},\theta_{i, 2}^{f},...,\theta_{i, K}^{f}]\\
        \bm\theta^g &= [\theta_{ij, 1}^{g},\theta_{ij, 2}^{g},...,\theta_{ij, K}^{g}]
    \end{aligned}
\end{equation}
In this study, the drift and the diffusion terms are taken independent of each other. The dictionaries for drift and diffusion terms $\boldsymbol{L}^{f}$ and $\boldsymbol{L}^{g}$ can be same or different. As the identification of the drift and diffusion terms is independent, two separate sparse learning frameworks one for the drift and another for the diffusion. Eq. \eqref{eq:lib} can be represented compactly by adding residual error vectors  $\boldsymbol{\varepsilon}_{i}$ and $\boldsymbol{\eta}_{i j}$ in drift and diffusion terms respectively as follows:
\begin{equation} \label{eq:compact}
    \begin{aligned}
        Y_i(t) &= f_i(\bm{X}(t)) + {\varepsilon}_{i}(t)\\
        &\approx \mathbf{L}^{f}(\bm{X}(t)) \boldsymbol{\theta}_{i}^{f}+{\varepsilon}_{i}(t)
    \end{aligned}
\end{equation}
upon discretizing, $t \rightarrow 1,2, \ldots ,K, \ldots, N$ we get,
\begin{equation} \label{eq:compact1}
    \begin{aligned}
    &\boldsymbol{Y}_{i}=\mathbf{L}^{f} \boldsymbol{\theta}_{i}^{f}+\boldsymbol{\varepsilon}_{i}\\
    \end{aligned}
\end{equation}
Similarly we can write the equation for diffusion as,
\begin{equation} \label{eq:compact2}
    \begin{aligned}
    &\boldsymbol{Y}_{i j}=\mathbf{L}^{g} \boldsymbol{\theta}_{i j}^{g}+\boldsymbol{\eta}_{i j}
    \end{aligned}
\end{equation}
where ${Y}_{i}={f}_{i}\left(\boldsymbol{X}(t), t\right)$ and ${Y}_{i j}={\Gamma}_{i j}\left(\boldsymbol{X}(t), t\right)$ are the target vectors of the sparse regression problem associated with $i^{t h}$-drift term and $(i j)^{t h}$-diffusion covariance term, respectively. Using sparse Bayesian linear regression, the above equations are solved to select the structure and obtain the parameter posterior over $\theta^f$ and $\theta^g$. To understand sparse Bayesian linear regression, let us represent Eq. \eqref{eq:compact1} and Eq. \eqref{eq:compact2} as follows:
\begin{equation}
\boldsymbol{Y}=\mathbf{L} \boldsymbol{\theta}+\bm{\epsilon}
\end{equation}
where $\boldsymbol{Y} \in \mathbb{R}^{N}$ denotes the $N$-dimensional target vector, $\mathbf{L}$ denotes the dictionary of basis functions, $\boldsymbol{\theta}$ is the weight vector. In case of drift identification $\mathbf{L}$ and $\boldsymbol{\theta}$ will be $\mathbf{L}^f$ and $\boldsymbol{\theta}^f$, Whereas for diffusion identification those will be $\mathbf{L}^g$ and $\boldsymbol{\theta}^g$. $\bm{\epsilon} \in \mathbb{R}^{N}$ is the residual error vector which represents the measurement error. Applying the Bayes formula, we obtain:
\begin{equation}
P(\boldsymbol{\theta} \mid \boldsymbol{Y})=\frac{P(\boldsymbol{\theta}) P(\boldsymbol{Y} \mid \boldsymbol{\theta})}{P(\boldsymbol{Y})}
\end{equation}
Where $P(\boldsymbol{\theta} \mid \boldsymbol{Y})$ is the posterior distribution of $\boldsymbol{\theta}$, $P(\boldsymbol{\theta})$ is the prior distribution, $P(\boldsymbol{Y} \mid \boldsymbol{\theta})$ is the likelihood function and $P(\boldsymbol{Y})$ is the marginal likelihood or evidence.
Modeling the measurement error $\bm{\epsilon}$ as i.i.d Gaussian random variable with zero mean and variance $\sigma^{2}$, the likelihood function is written as:
\begin{equation}
\boldsymbol{Y} \mid \boldsymbol{\theta}, \sigma^{2} \sim \mathcal{N}\left(\mathbf{L} \boldsymbol{\theta}, \sigma^{2} \mathbf{I}_{N \times N}\right)
\end{equation}
where $\mathbf{I}_{N \times N}$ denotes the $N$-dimensional identity matrix. As our aim is to discover the governing equations which represent the given dynamical system, it may be good to assume that the governing model will have only a few relevant terms from the dictionary. selecting such relevant terms may be achieved by using sparsity-promoting priors on the weight vector. In this work, we use spike and slab (SS) distribution to promote sparsity in the solution. The SS prior is a hierarchical discrete mixture prior, consisting of a Dirac-delta spike at zero and a continuous distribution. 
The SS prior promotes sparsity by classifying each component of the weight vector into either the spike or the slab. The components falling in the spike take zero values, whereas those falling in the slab can take non-zero values. This classification is controlled by an indicator variable $Z_k$ for each weight component '$k$'. If $Z_k$ takes a value of 1, the weight falls into the slab; else, it takes a value of 0 due to the spike. $\boldsymbol{\theta}_{r}$ is the group of weight vector which contains only those variables from $\boldsymbol{\theta}$ for which $Z_{k}=1$. The SS-prior is defined as:
\begin{equation}
p(\boldsymbol{\theta} \mid \boldsymbol{Z})=p_{s l a b}\left(\theta_{r}\right) \prod_{k, Z_{k}=0} p_{s p i k e}\left(\theta_{k}\right)
\end{equation}
where the spike and slab distributions are defined as, $p_{\text {spike }}\left(\theta_{k}\right)=\delta_{0}$ and $p_{\text {slab }}\left(\theta_{r}\right)=\mathcal{N}\left(\mathbf{0}, \sigma^{2} \vartheta_{s} \mathbf{R}_{0, r}\right)$ with $\mathbf{R}_{0, r}=\mathbf{I}_{r \times r}$. The hyperparameters $ p, \vartheta_{s}, a_{\sigma}$, and $b_{\sigma}$ in Fig. 1 are provided as deterministic constants in the hierarchical model. The random variables $\sigma^{2}, \boldsymbol{Z}$ and $\boldsymbol{\theta}$ are as follows:
\begin{equation}\label{bern}
    \begin{aligned}
    p(Z_{k} \mid p_{0}) &=\operatorname{Bern}
    (p_{0}) ; k=1 \ldots K \\
    p(\sigma^{2}) &=I G(\alpha_{\sigma}, \beta_{\sigma})
    \end{aligned}
\end{equation}
From Fig. 1, the joint distribution of the random variables $\boldsymbol{\theta}, \boldsymbol{Z}$ and $\sigma^{2}$ is obtained as, 
\begin{equation} \label{eq:Bayes}
\begin{aligned}
p(\boldsymbol{\theta}, \boldsymbol{Z}, \sigma^{2} \mid \boldsymbol{Y}) &=\frac{p(\boldsymbol{Y} \mid \boldsymbol{\theta}, \sigma^{2}) p(\boldsymbol{\theta} \mid \boldsymbol{Z}, \sigma^{2}) p(\boldsymbol{Z}) p(\sigma^{2})}{p(\boldsymbol{Y})} \\
& \propto p(\boldsymbol{Y} \mid \boldsymbol{\theta}, \sigma^{2}) p(\boldsymbol{\theta} \mid \boldsymbol{Z}, \sigma^{2}) p(\boldsymbol{Z}) p(\sigma^{2})
\end{aligned}
\end{equation}
where $p\left(\boldsymbol{\theta}, \boldsymbol{Z}, \sigma^{2} \mid \boldsymbol{Y}\right)$ denotes the joint distribution of the random variables, $p\left(\boldsymbol{Y} \mid \boldsymbol{\theta}, \sigma^{2}\right)$ denotes the likelihood function, $p\left(\boldsymbol{\theta} \mid \boldsymbol{Z}, \sigma^{2}\right)$ is the prior distribution for the weight vector $\boldsymbol{\theta}, p(\boldsymbol{Z})$ is the prior distribution for the latent vector $\boldsymbol{Z}, p\left(\sigma^{2}\right)$ is the prior distribution for the noise variance and $p(\boldsymbol{Y})$ is the marginal likelihood or evidence.

\subsection{Variational Bayesian inference for variable selection}
The Bayesian variable selection can be done using the SS priors from the posterior distribution $p\left(\boldsymbol{\theta}, \boldsymbol{Z}, \sigma^{2} \mid \boldsymbol{Y}\right)$ which can be computed using Bayes formula as shown in the Eq. \eqref{eq:Bayes}.
Unfortunately, it is not possible to compute the value of posterior analytically due to the presence of $p(\boldsymbol{Y})$ term, which entails a multi-dimensional intractable integral. MCMC-based methods give fairly accurate results, but they are computationally expensive. In this work, variational Bayes is used for approximating the joint posterior distribution  $p\left(\boldsymbol{\theta}, \boldsymbol{Z}, \sigma^{2} \mid \boldsymbol{Y}\right)$ by simpler variational distributions. However, it can be noticed that there is a Dirac-delta function in the SS prior which makes the derivation of the VB algorithm difficult. Therefore the linear regression model with SS prior needs to be reparameterised in a form that is more
suitable to the variational Bayes method \cite{nayek2021spike}. The SS prior is rewritten as 
\begin{equation}
    \begin{aligned}
    \boldsymbol{Y} \mid \boldsymbol{\theta}, \boldsymbol{Z}, \sigma^{2} &\sim \mathcal{N}\left(\mathbf{L} \boldsymbol{\Lambda} \boldsymbol{\theta}, \sigma^{2} \mathbf{I}_{N}\right), \\
    \sigma^{2} &\sim \mathcal{I} \mathcal{G}\left(a_{\sigma}, b_{\sigma}\right) \\
    \theta_{k} &\sim \mathcal{N}\left(0, \sigma^{2} v_{s}\right), \\
    Z_{k} &\sim \operatorname{Bern}\left(p_{0}\right), i=1, \ldots, K
    \end{aligned}
\end{equation}
where the term $\Lambda$ represents $\operatorname{diag}\left(Z_{1}, \ldots, Z_{K}\right)$. Variational Bayes approximates the true posterior distribution $p\left(\boldsymbol{\theta}, \boldsymbol{Z}, \sigma^{2} \mid \boldsymbol{Y}\right)$ by some probability distribution $q(\boldsymbol{\theta}, \boldsymbol{Z}, \sigma^{2})$ which belongs to some tractable family of distributions $Q$ such as Gaussian distribution. Thereafter, to find the best approximation $q^*\in Q$, the Kullback-Leibler (KL) divergence \cite{joyce2011kullback} between variational approximation $q(\boldsymbol{\theta}, \boldsymbol{Z}, \sigma^{2})$ and the true posterior $p\left(\boldsymbol{\theta}, \boldsymbol{Z}, \sigma^{2} \mid \boldsymbol{Y}\right)$ is minimized. The KL-divergence between the two distributions can be mathematically represented as
\begin{align}
q^{*}\left(\boldsymbol{\theta}, \boldsymbol{Z}, \sigma^{2}\right)
&=\underset{q \in Q}{\arg \min } \operatorname{KL}\left[q\left(\boldsymbol{\theta}, \boldsymbol{Z}, \sigma^{2}\right) \| p\left(\boldsymbol{\theta}, \boldsymbol{Z}, \sigma^{2} \mid \boldsymbol{Y}\right)\right]\\
&=\underset{q \in Q}{\arg \min }\label{eq:VB} \mathbb{E}_{q\left(\boldsymbol{\theta}, \boldsymbol{Z}, \sigma^{2}\right)}\left[\ln \left(\frac{q\left(\boldsymbol{\theta}, \boldsymbol{Z}, \sigma^{2}\right)}{p\left(\boldsymbol{\theta}, \boldsymbol{Z}, \sigma^{2} \mid \boldsymbol{Y}\right)}\right)\right]
\end{align}
where $\mathbb{E}_{q\left(\boldsymbol{\theta}, \boldsymbol{Z}, \sigma^{2}\right)}[\cdot]$ denotes the expectation with respect to the variational distribution $q\left(\boldsymbol{\theta}, \boldsymbol{Z}, \sigma^{2}\right)$. When Eq. \eqref{eq:VB} is expanded, a new term referred to as evidence lower bound (ELBO) is introduced, which plays an important role in assessing the convergence of the VB algorithm.
\begin{equation}
    \begin{aligned}
    &\mathrm{KL}\left[q\left(\boldsymbol{\theta}, \boldsymbol{Z}, \sigma^{2}\right) \| p\left(\boldsymbol{\theta}, \boldsymbol{Z}, \sigma^{2} \mid \boldsymbol{Y}\right)\right] \\ &=\mathbb{E}_{q\left(\boldsymbol{\theta}, \boldsymbol{Z}, \sigma^{2}\right)}\left[\ln \left(\frac{q\left(\boldsymbol{\theta}, \boldsymbol{Z}, \sigma^{2}\right)}{p\left(\boldsymbol{\theta}, \boldsymbol{Z}, \sigma^{2} \mid \boldsymbol{Y}\right)}\right)\right] \\
    &=\mathbb{E}_{q\left(\boldsymbol{\theta}, \boldsymbol{Z}, \sigma^{2}\right)}\left[\ln q\left(\boldsymbol{\theta}, \boldsymbol{Z}, \sigma^{2}\right)\right]-\mathbb{E}_{q\left(\boldsymbol{\theta}, \boldsymbol{Z}, \sigma^{2}\right)}\left[\ln \left(p\left(\boldsymbol{Y} \mid \boldsymbol{\theta}, \boldsymbol{Z}, \sigma^{2}\right) p\left(\boldsymbol{\theta}, \boldsymbol{Z}, \sigma^{2}\right)\right)\right]+\ln p(\boldsymbol{Y}) \\
    &=\mathbb{E}_{q\left(\boldsymbol{\theta}, \boldsymbol{Z}, \sigma^{2}\right)}\left[\ln \left(\frac{q\left(\boldsymbol{\theta}, \boldsymbol{Z}, \sigma^{2}\right)}{p\left(\boldsymbol{\theta}, \boldsymbol{Z}, \sigma^{2}\right)}\right)\right]-\mathbb{E}_{q\left(\boldsymbol{\theta}, \boldsymbol{Z}, \sigma^{2}\right)}\left[\ln p\left(\boldsymbol{Y} \mid \boldsymbol{\theta}, \boldsymbol{Z}, \sigma^{2}\right)\right]+\ln p(\boldsymbol{Y}) \\
    &={\operatorname{KL}\left[q\left(\boldsymbol{\theta}, \boldsymbol{Z}, \sigma^{2}\right) \| p\left(\boldsymbol{\theta}, \boldsymbol{Z}, \sigma^{2}\right)\right]-\mathbb{E}_{q\left(\boldsymbol{\theta}, \boldsymbol{Z}, \sigma^{2}\right)}\left[\ln p\left(\boldsymbol{Y} \mid \boldsymbol{\theta}, \boldsymbol{Z}, \sigma^{2}\right)\right]}+\ln p(\boldsymbol{Y}) \\
    &=\ln p(\boldsymbol{Y})-\mathrm{ELBO}
    \end{aligned}
\end{equation}
where ${\operatorname{KL}\left[q\left(\boldsymbol{\theta}, \boldsymbol{Z}, \sigma^{2}\right) \| p\left(\boldsymbol{\theta}, \boldsymbol{Z}, \sigma^{2}\right)\right]-\mathbb{E}_{q\left(\boldsymbol{\theta}, \boldsymbol{Z}, \sigma^{2}\right)}\left[\ln p\left(\boldsymbol{Y} \mid \boldsymbol{\theta}, \boldsymbol{Z}, \sigma^{2}\right)\right]} $ = $-$ELBO. The term $\ln p(\boldsymbol{y})$ is constant with respect to the distribution $q\left(\boldsymbol{\theta}, \boldsymbol{Z}, \sigma^{2}\right)$. As KL divergence is a non-negative quantity, the ELBO can be seen as the lower bound to $\ln p(\boldsymbol{Y})$, therefore minimizing KL divergence, $\mathrm{KL}\left[q\left(\boldsymbol{\theta}, \boldsymbol{Z}, \sigma^{2}\right) \| p\left(\boldsymbol{\theta}, \boldsymbol{Z}, \sigma^{2} \mid \boldsymbol{Y}\right)\right]$ is equivalent to maximizing the ELBO, thus,
\begin{equation} \label{eq:ELBO}
q^{*}\left(\boldsymbol{\theta}, \boldsymbol{Z}, \sigma^{2}\right)=\underset{q \in Q}{\arg \max } \underbrace{\mathbb{E}_{q\left(\boldsymbol{\theta}, \boldsymbol{Z}, \sigma^{2}\right)}\left[\ln p\left(\boldsymbol{Y} \mid \boldsymbol{\theta}, \boldsymbol{Z}, \sigma^{2}\right)\right]-\mathrm{KL}\left[q\left(\boldsymbol{\theta}, \boldsymbol{Z}, \sigma^{2}\right) \| p\left(\boldsymbol{\theta}, \boldsymbol{Z}, \sigma^{2}\right)\right]}_{\text {ELBO }}
\end{equation}
In this work, $q\left(\boldsymbol{\theta}, \boldsymbol{Z}, \sigma^{2}\right)$ has been chosen to have the following factorized form,
\begin{equation}
q\left(\boldsymbol{\theta}, \boldsymbol{Z}, \sigma^{2}\right)=q(\boldsymbol{\theta}) q\left(\sigma^{2}\right) \prod_{i=1}^{K} q\left(Z_{i}\right) 
\end{equation}
and the corresponding individual variational distributions are selected as,
\begin{align}
    q(\boldsymbol{\theta}) &=\mathcal{N}\left(\boldsymbol{\mu}^{q}, \boldsymbol{\Sigma}^{q}\right) \\
    q\left(\sigma^{2}\right) &=\mathcal{I} G\left(a_{\sigma}^{q}, b_{\sigma}^{q}\right) \\
    q\left(Z_{k}\right) &=\operatorname{Bern}\left(w_{k}^{q}\right), \text { for } i=1, \ldots, K
\end{align}
Here, $\boldsymbol{\mu}^{q}, \boldsymbol{\Sigma}^{q}, a_{\sigma}^{q}, b_{\sigma}^{q}, w_{i}^{q}$ represent the deterministic variational parameters. Values of these parameters need to be optimized to minimize the KL divergence between the approximate variational distribution and the true posterior distribution (see Eq.\eqref{eq:VB}). The optimal choice of variational parameters that maximize the ELBO in Eq.\eqref{eq:ELBO} are found to satisfy the following relations,
\begin{align}
q^{*}(\boldsymbol{\theta}) & \propto \mathbb{E}_{q(\boldsymbol{Z}) q\left(\sigma^{2}\right)}\left[\ln p\left(\boldsymbol{Y}, \boldsymbol{\theta}, \boldsymbol{Z}, \sigma^{2}\right)\right] \\
q^{*}(\boldsymbol{Z}) & \propto \mathbb{E}_{q(\boldsymbol{\theta}) q\left(\sigma^{2}\right)}\left[\ln p\left(\boldsymbol{Y}, \boldsymbol{\theta}, \boldsymbol{Z}, \sigma^{2}\right)\right] \\
q^{*}\left(\sigma^{2}\right) & \propto \mathbb{E}_{q(\boldsymbol{\theta}) q(\boldsymbol{Z})}\left[\ln p\left(\boldsymbol{Y}, \boldsymbol{\theta}, \boldsymbol{Z}, \sigma^{2}\right)\right]
\end{align}
and when the above equations are solved, the following expressions for the variational parameters are obtained \cite{nayek2022equation}:
\begin{subequations}\label{eq:variationa_param}
    \begin{align}
        \boldsymbol{\Sigma}^{q} &=\left[\tau\left(\left(\mathbf{L}^{T} \mathbf{L}\right) \odot \boldsymbol{\Omega}+v_{s}^{-1} \mathbf{I}_{P}\right)\right]^{-1} \\
        \boldsymbol{\mu}^{q} &=\tau \boldsymbol{\Sigma}^{q} \mathbf{W}^{q} \mathbf{L}^{T} \boldsymbol{Y} \\
        a_{\sigma}^{q} &=a_{\sigma}+0.5 N+0.5 P \\
        b_{\sigma}^{q} &=b_{\sigma}+0.5\left[\boldsymbol{Y}^{T} \boldsymbol{Y}-2 \boldsymbol{Y}^{T} \mathbf{L W}^{q} \boldsymbol{\mu}^{q}+\operatorname{tr}\left\{\left(\left(\mathbf{L}^{T} \mathbf{L}\right) \odot \boldsymbol{\Omega}+v_{s}^{-1} \mathbf{I}_{P}\right)\left(\boldsymbol{\mu}^{q} \boldsymbol{\mu}^{q T}+\boldsymbol{\Sigma}^{q}\right)\right\}\right], \\
        \tau &=a_{\sigma}^{q} / b_{\sigma}^{q}, \\
        \eta_{k} &=\operatorname{logit}\left(p_{0}\right)-0.5 \tau\left(\left(\mu_{k}^{q}\right)^{2}+\Sigma_{k, k}^{q}\right) \boldsymbol{h}_{k}^{T} \boldsymbol{h}_{k}+\tau \boldsymbol{h}_{k}^{T}\left[\boldsymbol{Y} \mu_{k}^{q}-\mathbf{L}_{-k} \mathbf{W}_{-k}^{q}\left(\boldsymbol{\mu}_{-k}^{q} \mu_{k}^{q}+\boldsymbol{\Sigma}_{-k, k}^{q}\right)\right] \\
        w_{k}^{q} &=\operatorname{expit}\left(\eta_{k}\right)
    \end{align}
\end{subequations}
In the above expressions, $\operatorname{logit}(A)=\ln (A)-\ln (1-A), \operatorname{expit}(A)=\operatorname{logit}^{-1}(A)=\exp (A) /(1+\exp (A)), \boldsymbol{w}^{q}=$ $\left[w_{1}^{q}, \ldots, w_{K}^{q}\right]^{T}, \mathbf{W}^{q}=\operatorname{diag}\left(\boldsymbol{w}^{q}\right), \boldsymbol{\Omega}=\boldsymbol{w}^{q} \boldsymbol{w}^{q T}+\mathbf{W}^{q}\left(\mathbf{I}_{K}-\mathbf{W}^{q}\right)$ and the symbol $\odot$ denotes the element-wise multiplication between two matrices, $\boldsymbol{h}_{i}$ denotes the $i^{\text {th }}$ column of $\mathbf{L}$, whereas the notation $\mathbf{L}_{-i}$ represents the dictionary matrix with the $i^{\text {th }}$ column removed. One can observe that variational parameters do not have explicit solutions, instead, their update expressions are dependent upon each other. So to optimize these parameters, an iterative coordinate-wise updating method is followed. In this procedure, the variational parameters are first initialized and then cyclically updated conditioned upon the updates of other parameters in the most recent iteration.\\\
For the commencement of the VB algorithm, the deterministic parameters are set to the following values: a slab variance $v_{s}=10$, noise prior parameters $a_{\sigma}=10^{-4}, b_{\sigma}=10^{-4}$, and a small probability $p_{0}=0.1$, which favors the selection of simpler models. Initialization of the variational parameters $\left\{\boldsymbol{w}^{q}, \tau\right\}$ is also a vital task as the algorithm is found to be quite sensitive to the initial choice of $\boldsymbol{w}^{q}$ which represents the vector of inclusion probabilities of the basis variables. In this work, the Sparse Bayesian Learning (SBL) method \cite{tipping1999relevance} has been used to initialize $\boldsymbol{w}^{q}$. More specifically, the initial $\boldsymbol{w}^{q}$ is set equal to a diagnostic parameter $\gamma$ outputted from SBL. The SBL diagnostic parameter $\gamma_{i} \in(0,1)$ represents a probability measure of how important the $i^{th}$  column of $\bm{L}$ is in explaining the target vector $\boldsymbol{Y}$. Moreover, an SBL run is very cheap and requires very little time to get the initial guess $\boldsymbol{w}^{(0)}$. The initial value of $\tau$, which represents the expected precision of the noise $\sigma^2$, is set to 1000.\\\
The VB algorithm iteratively maximizes the ELBO until it converges to a local maximum of the bound. The variational parameters are updated in every iteration, starting with the initial values until convergence. The convergence criteria here is set as the difference between the ELBO value of two successive iterations. When it reaches a small value $\rho$ (say equal to $10^{-6}$), updating is stopped, and the parameters are taken as optimized variational parameters.
\begin{equation}
\operatorname{ELBO}^{(t)}-\operatorname{ELBO}^{(t-1)}<\rho
\end{equation}
The value of ELBO, at each iteration $t$, is computed using the simplified expression:
\begin{equation}\label{eq:elbo}
    \begin{aligned}
        \mathrm{ELBO}^{(t)} &=0.5 P-0.5 N \ln (2 \pi)-0.5 P \ln \left(v_{s}\right)+a_{\sigma} \ln \left(b_{\sigma}\right)-\ln \Gamma\left(a_{\sigma}\right)+\ln \Gamma\left(a_{\sigma}^{(t)}\right)\\
        &-a_{\sigma}^{(t)} \ln \Gamma\left(b_{\sigma}^{(t)}\right)
        +0.5 \ln \left|\boldsymbol{\Sigma}^{(t)}\right|+\sum_{i=1}^{K}\left[w_{i}^{(t)} \ln \left(\frac{p_{0}}{w_{i}^{(t)}}\right)+\left(1-w_{i}^{(t)}\right) \ln \left(\frac{1-p_{0}}{1-w_{i}^{(t)}}\right)\right]
    \end{aligned}
\end{equation}
where $\Gamma(\cdot)$ is the Gamma function, and $a_{\sigma}^{(t)}, b_{\sigma}^{(t)}, \boldsymbol{\mu}^{(t)}, \boldsymbol{\Sigma}^{(t)}, \boldsymbol{w}^{(t)}$, denote the variational parameters at the $t^{\text {th }}$ iteration, dropping the '$q$' superscript. Upon convergence, the variational parameters from the final step are denoted by $a_{\sigma}^{*}, b_{\sigma}^{*}, \mu^{*}, \Sigma^{*}, \boldsymbol{w}^{*}$.\\\
For all demonstrations in this work, the data for the SDE is generated using the Euler-Maruyama scheme with the sampling frequency set as 1000 Hz. The candidate functions that are considered inside the dictionary $\mathbf{L} \in \mathbb{R}^{N \times K}$ are given below. The function represents a mapping of the $m$-dimensional state vector $\boldsymbol{X}=\left\{X_{1}, X_{2}, \ldots X_{m}\right\}$.
\begin{equation}\label{eq:library}
	{\bf{L}}({\bm{X}}) = \left[ \begin{array}{*{20}{c}}
	{\bf{1}}&{\bm{X}}&{P^\mathcal{P}}({\bm{X}})&{{\mathop{\rm sgn}} ({\bm{X}})}&{\left| {\bm{X}} \right|}&{{\bm{X}}\left| {\bm{X}} \right|}&{{\mathop{\rm sin}} ({\bm{X}})}&{{\mathop{\rm cos}} ({\bm{X}})}
	\end{array} \right]
\end{equation}
Here, $\mathbf{1} \in \mathbb{R}^{N}$ denotes an $N$-dimensional vector of ones, and $P^{\mathcal{P}}(\cdot) \in \mathbb{R}^{N \times m}$ denotes a set of terms present in the multinomial expansion $\left(X_{1}+\ldots+X_{m}\right)^{\mathcal{P}}$.
The functions $\operatorname{sgn}(.) \in \mathbb{R}^{N \times m}$, and $|(\cdot)| \in \mathbb{R}^{N \times m}$, $(\cdot)|(\cdot)| \in \mathbb{R}^{N \times 2 m}$, $\operatorname{sin}(.) \in \mathbb{R}^{N \times m}$, and $\operatorname{cos}(.) \in \mathbb{R}^{N \times m}$ are the signum, absolute, and tensor product, sine and cosine functions of the state vector.
The order of the polynomial $\mathcal{P}$ is chosen to be different for different numerical problems depending upon the highest degree of the polynomial contained in the equation to be discovered.
The final model is selected on the basis of marginal posterior inclusion probability(PIP), $p\left(Z_{i}=1 \mid \boldsymbol{Y}\right)$.
Those basis functions which have greater than half the PIP value are selected in the final model. In VB inference, the estimated $w_{i}^{*}$ s can be interpreted as an approximation to the posterior probability of $p\left(Z_{i}=1 \mid \boldsymbol{Y}\right)$. Post inference, the basis variables which have $w_{i}^{*}>0.5$ are included in the final estimated model. The estimated mean and covariance of the weight vectors $\boldsymbol{\theta}$, denoted by $\hat{\boldsymbol{\mu}}_{\boldsymbol{\theta}}$ and $\hat{\boldsymbol{\Sigma}}_{\boldsymbol{\theta}}$, are respectively populated with values of $\boldsymbol{\mu}^{*}$ and $\boldsymbol{\Sigma}^{*}$ at the respective indices which correspond to the selected basis variables, and the remaining entries of $\hat{\boldsymbol{\mu}}_{\boldsymbol{\theta}}$ and $\hat{\boldsymbol{\Sigma}}_{\boldsymbol{\theta}}$ are set to zero. Thereafter, using the mean and covariance of the weights, the predictions with the estimated model can be made as shown below,
\begin{align}
\boldsymbol{\mu}_{\boldsymbol{Y}^{*}} &=\mathbf{L}^{*} \hat{\boldsymbol{\mu}}_{\boldsymbol{\theta}} \\
\mathbf{V}_{\boldsymbol{Y}^{*}} &=\mathbf{L}^{*} \hat{\boldsymbol{\Sigma}}_{\boldsymbol{\theta}} \mathbf{L}^{* T}+\left(a_{\sigma}^{*} / b_{\sigma}^{*}\right)^{-1} \mathbf{I}_{N^{*}}
\end{align}
where $\mathbf{L}^{*}$ is the $N^{*} \times P$ test dictionary, defined at a set of $N^{*}$ previously-unseen test data points, $\boldsymbol{\mu}_{\boldsymbol{Y}^{*}}$ is the $N^{*} \times 1$ predicted mean of the target vector and $\mathbf{V}_{\boldsymbol{Y}^{*}}$ is the predicted covariance of the target vector.\\\
While constructing the dictionary, it is observed that the dictionary is often ill-conditioned. This mainly happens due to the combined effect of the large-scale difference between the basis variables and the strong linear correlation between certain basis variables. So, to reduce the effect of these problems, the dictionary needs to be standardized. Here, the columns of the dictionary are centered and scaled to have zero mean and unit standard deviation. In addition to that, the target vector is set to have zero mean. The standardized dictionary and the target vector can mathematically be represented in the following form:
\begin{align}\label{eq:standardize}
    &\mathbf{L^{s}} = ( \mathbf{L} - \mathbf{1}{\boldsymbol{\mu}_{\mathbf{D}}){\mathbf{S}_{\mathbf{D}}}^{-1}}\\
    &\boldsymbol{Y}^{s} = \boldsymbol{Y} - \mathbf{1}\boldsymbol{\mu}_{\boldsymbol{Y}}
\end{align}
where $\boldsymbol{1}$ denotes a column vector of ones, $\boldsymbol{\mu}_{\mathbf{D}}$ is a row vector of the column-wise means of the dictionary $\mathbf{L}$, ${S_{\mathbf{D}}}$ is a diagonal
matrix of the column-wise standard deviations of $\mathbf{L}$, and $\boldsymbol{\mu}_{\boldsymbol{Y}}$ is the mean of the target training vector ${\boldsymbol{Y}}$. To prevent ill-conditioning it is important to standardize the dictionary and scale the target vector, before performing the VB algorithm. As such, the weights obtained are scaled weights $\boldsymbol{\theta}^S$ after running the VB algorithm. Hence, the weights are needed to transform back to the original space using the relations:
$\boldsymbol{\hat{\mu}_{\theta}} = {\mathbf{S}_{\mathbf{D}}}^{-1} {\boldsymbol{\hat{\mu}_{\theta^s}}}$ and $\boldsymbol{\hat{\Sigma}_{\theta}} = {\mathbf{S}_{\mathbf{D}}}^{-1} {\boldsymbol{\hat{\Sigma}_{\theta^s}}} {\mathbf{S}_{\mathbf{D}}}^{-1} $
\begin{algorithm}[ht!]
	\caption{Pseudo-code of the Variational Bayesian Inference}\label{algvecvec}
	\begin{algorithmic}[1]
	\Require{Displacement measurements: ${\bf{X}}(t) \in \mathbb{R}^{N \times m}$ and hyperparameters: $ p, \vartheta_{s}, a_{\sigma}$, $b_{\sigma}$, error bound: $\rho$.}
	    \State{Construct the dictionary ${\bf{L}}$ using the candidate basis functions (${\bf{L}^f}$ for drift ${\bf{L}^g}$ for diffusion).}
	    \State{Set the initial value of $\tau$ and initialise the variational parameter $\boldsymbol{w}^{q}$ using RVM.}
	   \State{Standardize the dictionary $\mathbf{L}$ and scale the training target vector $\boldsymbol{Y}$.} \Comment{Eq.\eqref{eq:standardize}}
		\While {$\operatorname{ELBO}^{(t)}-\operatorname{ELBO}^{(t-1)}>\rho$}
		\State{Update the variational parameters $\boldsymbol{\Sigma}^{q}$, $\boldsymbol{\mu}^{q}$, $a_{\sigma}^{q}$, $b_{\sigma}^{q}$, $\tau,\eta_{i}$, and $w_{i}^{q}$.}\label{step1} \Comment{Eq.\eqref{eq:variationa_param}}
		\State {Calculate the ELBO using the variational parameters for the iteration- $t$}\label{stepn} \Comment{Eq.\eqref{eq:elbo}}
        \State{Repeat steps \ref{step1}$ \to $\ref{stepn}}
		\EndWhile
	\State {Transform the output into original space using $\boldsymbol{\hat{\mu}_{\theta}} = {\mathbf{S}_{\mathbf{D}}}^{-1} {\boldsymbol{\hat{\mu}_{\theta^s}}}$ and $\boldsymbol{\hat{\Sigma}_{\theta}} = {\mathbf{S}_{\mathbf{D}}}^{-1} {\boldsymbol{\hat{\Sigma}_{\theta^s}}} {\mathbf{S}_{\mathbf{D}}}^{-1} $}
	\State{Calculate the marginal PIP values $p(Z_k=1|{\bm{Y}})$.} \Comment{Eq.\eqref{bern}}
	\State{Select the basis function in the final model which have a greater PIP value than the threshold.}
	\Ensure{The mean ${\bm{\hat{\mu}}}_{\theta}$ and covariance ${\bf{\hat{\Sigma}}}_{\theta}$.}
	\end{algorithmic}
\end{algorithm}

\section{Numerical problems}\label{problems}
In this section, the efficacy and robustness of  MAntRA is illustrated with three numerical examples. For all the examples, noisy measurements of displacements for a period of one second, $\tau \in [0,1]$, and the objective is to compute the probability of failure at different time instants. In step 2 of MAntRA, we have employed the well-known Euler-Maruyama (E-M) schemes. The reference solutions (ground truth) are generated by using the original SDE and the E-M scheme. Other specific details regarding sampling frequency, number of realizations and time-step size are provided with each example separately.


\subsection{Example 1: SDOF Duffing oscillator}\label{sec:system1}
As the first example, we consider an SDOF Duffing oscillator excited using Gaussian white noise. Duffing oscillator is a second-order nonlinear dynamical system with cubic stiffness non-linearity. It often finds its application in the modeling of electrical circuits, ionization waves, beam buckling, nonlinear hardening/softening springs, and models of flow-induced structural vibration problems. The equation of the oscillator is expressed as, 
\begin{equation}
    m \ddot{X}(t)+c \dot{X}(t)+k X(t)+\alpha X^{3}(t)=\sigma \dot{B}(t); \quad X\left(t=t_{0}\right)={X}_{0} ; \quad t \in[0, T]
\end{equation}
where $m \in \mathbb{R}$, $c \in \mathbb{R}$, and $k \in \mathbb{R}$ are the mass, damping, and stiffness parameters of the oscillator, $\alpha \in \mathbb{R}$ is a parameter associated with the cubic non-linearity, $\sigma$ is a non-negative parameter representing the strength of the additive white noise, and $B(t)$ is the Brownian motion. Here the derivative of the Brownian motion $\dot{B}(t)$ represents the zero mean Gaussian white noise. The model parameters considered in this example are shown in the second column of Table \ref{tab:tab1}.

\begin{table}[ht]
    \centering
    \begin{tabular}{c|c|c}
    \hline
    Parameter & Actual value & Predicted value\\
    \hline
    Damping (Ns/m) & $c/m$ = $2.00$ & $c/m$ = $2.00$\\
    Stiffness (N/m) & $k/m$ = $1000.00$ & $k/m$ = $999.87$\\
    Diffusion  & $\sigma/m$ = $1.00$ & $\sigma/m$ = $1.09$\\
    Nonlinear & $\alpha/m$ = $100000.00$ & $\alpha/m$ = $100369.79$\\
    \hline
    \end{tabular}
    \caption{System parameters for Example problem 1 - SDOF Duffing oscillator}\label{tab:tab1}
\end{table}

Considering $\{X, \dot{X} \}$=$\{X_{1}, X_{2} \}$, the corresponding first-order It\^{o}-stochastic SDEs for the dynamical system can be expressed as:
\begin{equation}\label{eq:fo_duffing}
    \left[\begin{array}{l}
        d X_{1}(t) \\
        d X_{2}(t)
        \end{array}\right]=\left[\begin{array}{c}
        X_{2}(t) \\
        -\frac{1}{m}\left(k X_{1}(t)+c X_{2}(t)+\alpha X_{1}^{3}(t)\right)
        \end{array}\right] d t+\left[\begin{array}{c}
        0 \\
        \frac{\sigma}{m} X(t)
    \end{array}\right] d B(t)   
\end{equation}
To illustrate the performance of the proposed MAntRA, we have generated synthetic data by solving Eq. \eqref{eq:fo_duffing} using E-M scheme. A time-step of {$\Delta t = 0.001$} is considered and, as previously stated, data corresponding to $t\in[0,1]$s is generated. To track the aleatoric uncertainty, {$N_t = 500$} such realizations were generated. The generated realizations were corrupted with {$5\%$} noise.

\begin{figure}[ht!]
    \centering
    \includegraphics[width=0.75\textwidth]{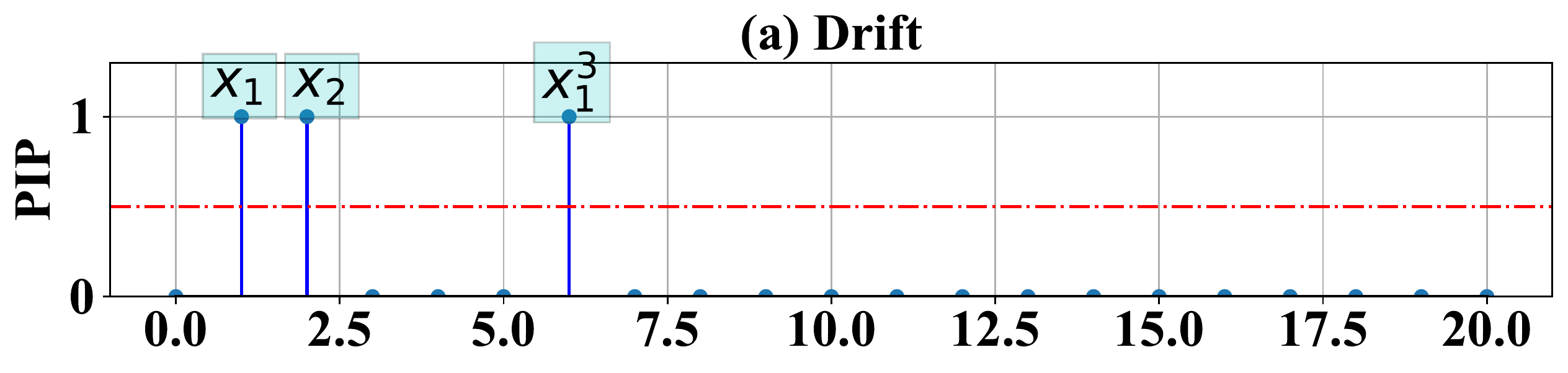}
    \includegraphics[width=0.75\textwidth]{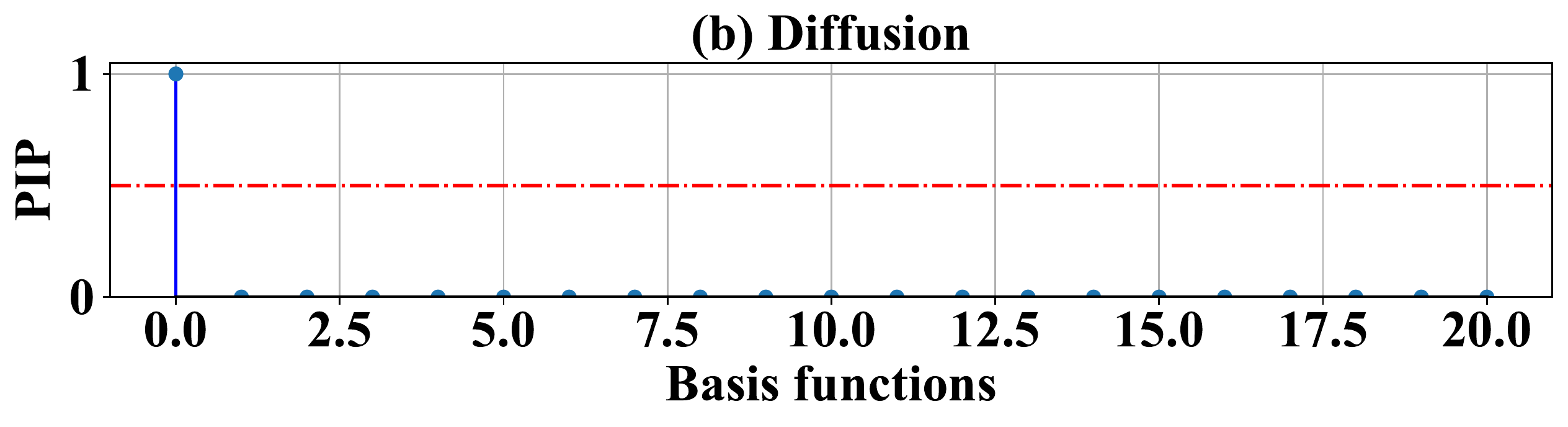}
    \caption{\textbf{Basis function selection for an SDOF Duffing oscillator system.} The dictionary $\textbf{L} \in \mathbb{R}^{N\times21}$ of 21 basis functions is used, which contains polynomial terms upto fourth order. The basis functions representing the equation are picked by the model with PIP almost equal to 1, (a)Basis functions selection for drift term. Functions that are picked by the model are $<X_1$, $X_2$, $X_1^3>$, (b) Basis functions selection for diffusion term. The predicted function in the second equation is a constant.}\label{ddr1}
\end{figure}

The proposed MAntRA was employed based on the synthetic data discussed above. However, before proceeding with the discussion on the reliability estimates obtained using the proposed approach, we discuss the performance of the proposed variational Bayesian equation discovery algorithm, which is an integral part of the proposed MAntRA. Fig. \ref{ddr1} shows the identified basis function corresponding to the drift and diffusion components. It is evident that the proposed framework is able to accurately discover the basis functions for the drift and diffusion terms as, $\left\{X_{1}(t), X_{2}(t), X_{1}^{3}(t)\right\}$ and constants respectively. The corresponding parameters are shown in Table \ref{tab:tab1} (last column). Again, the estimated parameters match almost exactly with the actual values. This clearly indicates that the developed variational Bayesian SDE discovery framework is capable of identifying the governing SDE from  one second of noisy displacement measurements, sampled at a sampling frequency of $1000$Hz.

\begin{figure}[!ht]
    \centering
    \includegraphics[width=0.75\textwidth]{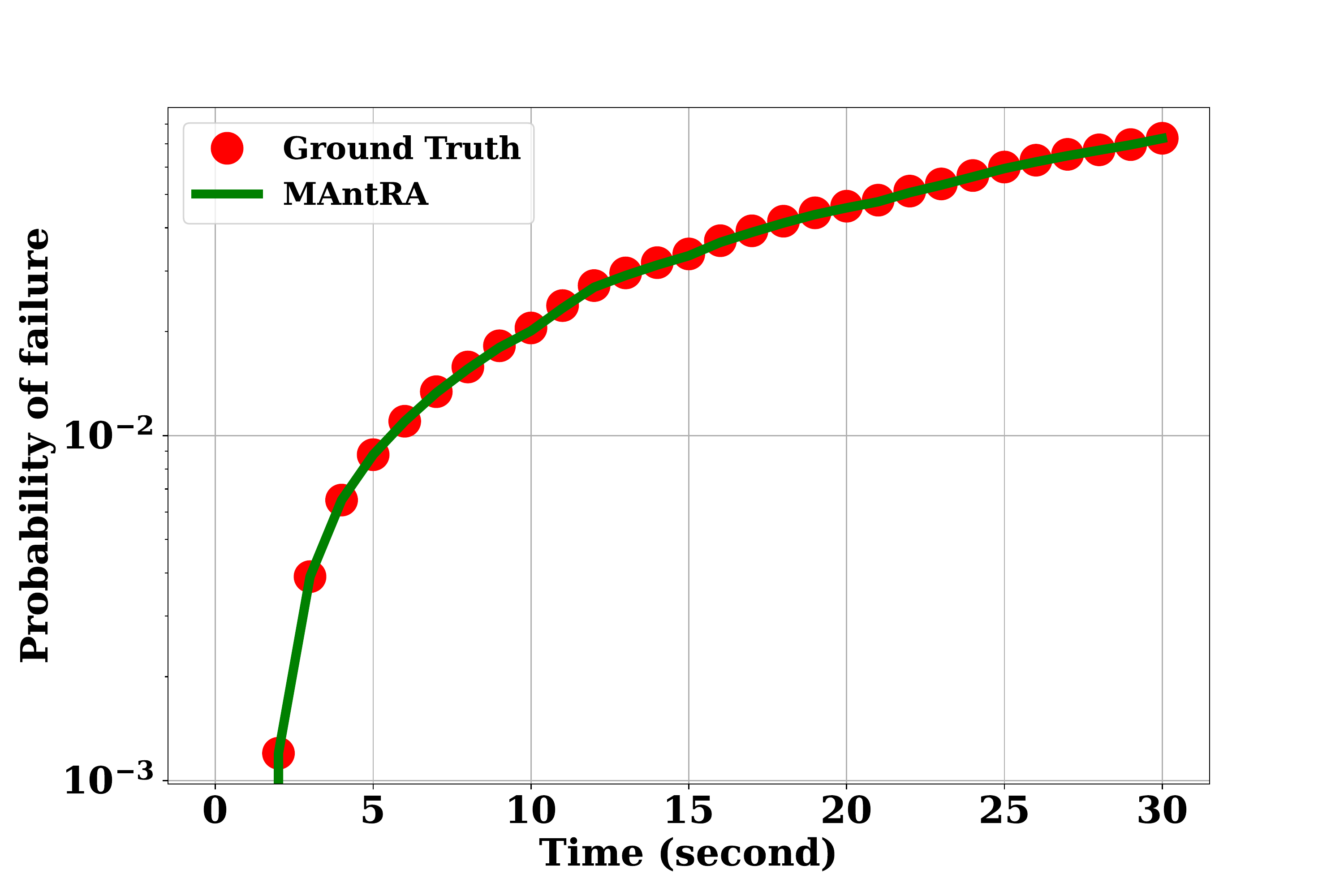}
    \caption{Probability of failure of an SDOF Duffing oscillator system using true and discovered models}
    \label{fig:rel_duff}
\end{figure}

Having validated the equation discovery module of the proposed MAntRA, we proceed with the performance of the proposed MAntRA is solving time-dependent reliability analysis problem. To that end, a limit-state function $\mathcal J \left(\zeta \right)$ is defined as follows:
\begin{equation}\label{eq:limit-state1}
    \mathcal J \left( \zeta, t \right) = X(\zeta,t) - X_t,
\end{equation}
where $X(\zeta,t)$ is the predicted response at time $t$ and $X_t$ is the threshold. $X_t = 0.0645$ is considered in this study. In a general settings, the threshold $X_t$ can be a function of time $t$ as well; however, in this paper, we have considered it to be a constant. Using the limit-state function $\mathcal J \left( \zeta, t \right)$, the probability of failure $P_f(t)$ at a time $t$ is computed as follows:
\begin{equation}\label{eq:pf}
    P_f (t) = \int_{\Omega_f} \int_0^t  d\tau d\zeta,
\end{equation}
where $\Omega_f$ represents the failure domain. Fig. \ref{fig:rel_duff} shows the probability of failure at different time-instants. We observe that the results obtained using the proposed MAntRA match exactly with the reference solution obtained by using the true SDE. More importantly, MAntRA yields excellent result even at $t=30$s, which is at a distant location from the training window.



\subsection{3-DOF nonlinear oscillator}
In the second example, we considered a 3-DOF nonlinear oscillator. The non-linearity in the oscillator is modeled using springs with cubic restoring forces. The motion equations of the oscillator can be represented in the form of the spring-mass-dashpot model as follows:
\begin{equation}\label{eq:3dof}
    \textbf{M}\bm{\ddot{X}}(t)+ \textbf{C}\bm{\dot{X}}(t)+ \textbf{K}\bm{X}(t)+ \bm{\Tilde{N}}(\bm{X},\dot{\bm{X}},t) = \bm{\Sigma}\bm{\dot{B}}(t)
\end{equation}
where $\mathbf{M} \in \mathbb{R}^{3 \times 3}$ is the mass matrix, $\mathbf{C} \in \mathbb{R}^{3 \times 3}$ is the damping matrix, $\mathbf{K} \in \mathbb{R}^{3 \times 3}$ is the stiffness matrix, $\bm{\dot{B}}(t) \in \mathbb{R}^{3}$ is the vector containing the derivatives of the Brownian motion, and $\bm{\Sigma}: \mathbb{R}^{3} \mapsto \mathbb{R}^{3 \times 3}$ is the matrix containing the strength of the additive white noise. The term $\bm{\Tilde{N}}(\cdot): \mathbb{R}^{3} \mapsto \mathbb{R}^{3}$ represents the nonlinear components of the oscillator which is given as,
\begin{equation}\label{eq:smd}
    \bm{\Tilde{N}}(\bm{X},\dot{\bm{X}},t) = \left[\begin{array}{c}
    \alpha_1 X_1^3 + \alpha_2 (X_1 - X_3)^3 \\
    \alpha_2 (X_3 - X_1)^3 + \alpha_3 (X_3 - X_5)^3 \\
    \alpha_3 (X_5 - X_3)^3 \\
    \end{array}\right]
\end{equation}
The state-space $\{ X_1, \dot{X}_1, X_2, \dot{X}_2, X_3, \dot{X_3} \}$ = $\{ X_1,X_{2},{X_3},X_{4},{X_5},X_{6} \}$ are used to construct the first-order It\^{o} SDEs for the system. The system parameters are given in Table \ref{tab:tab2}. Using the statespace one can find the corresponding first-order It\^{o} SDEs and arrange it in the form in Eq.\eqref{eq:goveq}. For data simulation, the initial conditions are taken as $\bm{X}(0)$ = [0.05, 0, 0.01, 0, 0.01, 0].

\begin{figure}[!ht]
    \centering
    \includegraphics[width=0.75\textwidth]{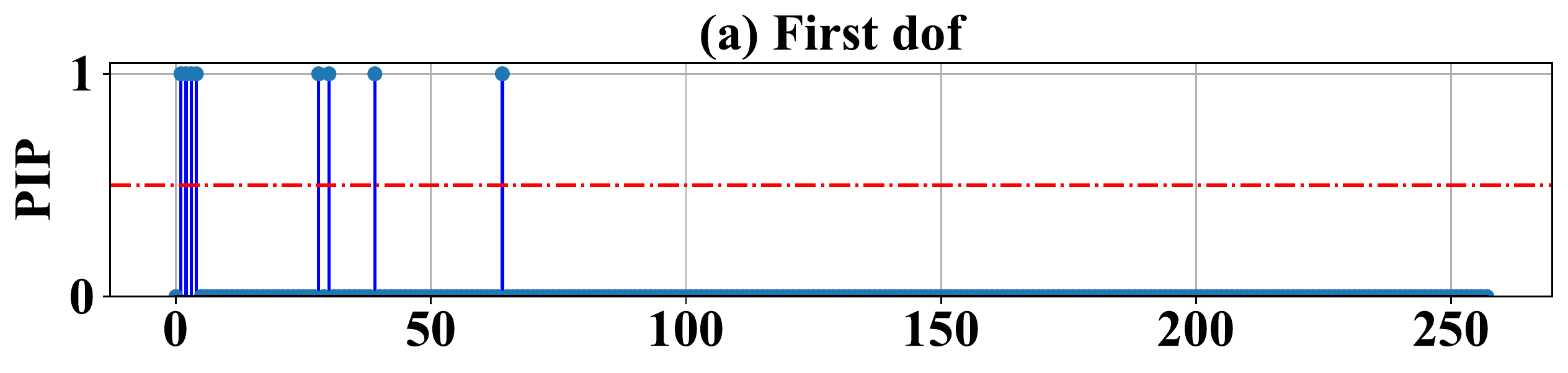}  
    \includegraphics[width=0.75\textwidth]{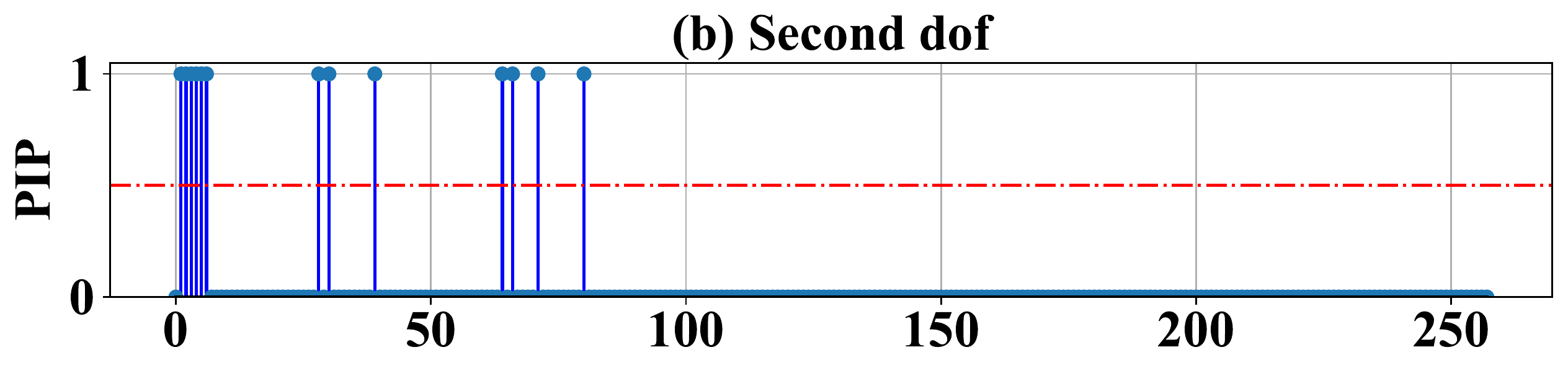}
    \includegraphics[width=0.75\textwidth]{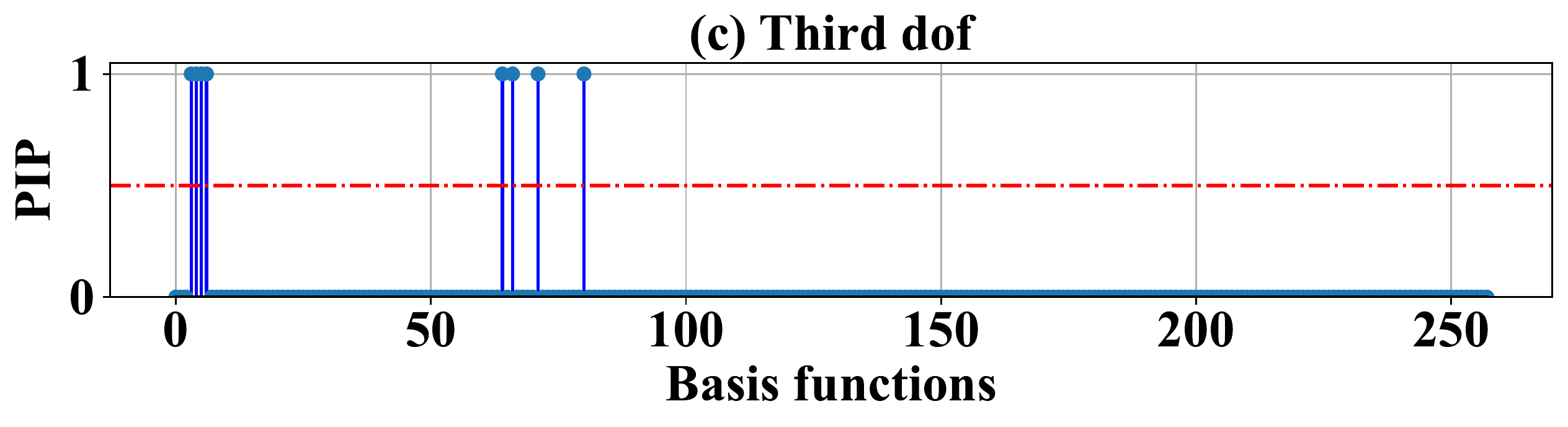}
    \caption{\textbf{Basis function selection for drift component of a 3-DOF Duffing oscillator system.} The dictionary $\textbf{L} \in \mathbb{R}^{N\times258}$ of 258 basis functions is used, which contains polynomial terms upto fourth order. The basis functions representing the equation are picked by the model with PIP almost equal to 1. (a) Basis functions selection for equation 1 of the drift term. The basis functions picked by the model for this equation are $<X_1$, $X_2$, $X_3$, $X_4$, $X_1^3$, $X_1^{2}X_3$, $X_1X_3^2$, $X_3^{3}>$. (b) Basis functions selection for equation 2 of drift term. The predicted basis functions in second equation are $<X_1$, $X_2$, $X_3$, $X_4$, $X_5$, $X_6$, $X_1^3$, $X_1^{2}X_3$, $X_1X_3^2$, $X_3^{3}$, $X_3^{2}X_5$, $X_5^{2}X_3$, $X_5^{3}>$. (c) Basis functions selection for equation 3 of drift term. The predicted basis functions in the third equation are $<X_3$, $X_4$, $X_5$, $X_6$, $X_3^{3}$, $X_3^{2}X_5$, $X_5^{2}X_3$, $X_5^{3}>$. The parameter values associated with the selected basis functions predicted by the model are shown in Table \ref{tab:tab2}}
    \label{fig:3dof_driftl}
\end{figure}

\begin{figure}[!ht]
    \centering
    \includegraphics[width=0.75\textwidth]{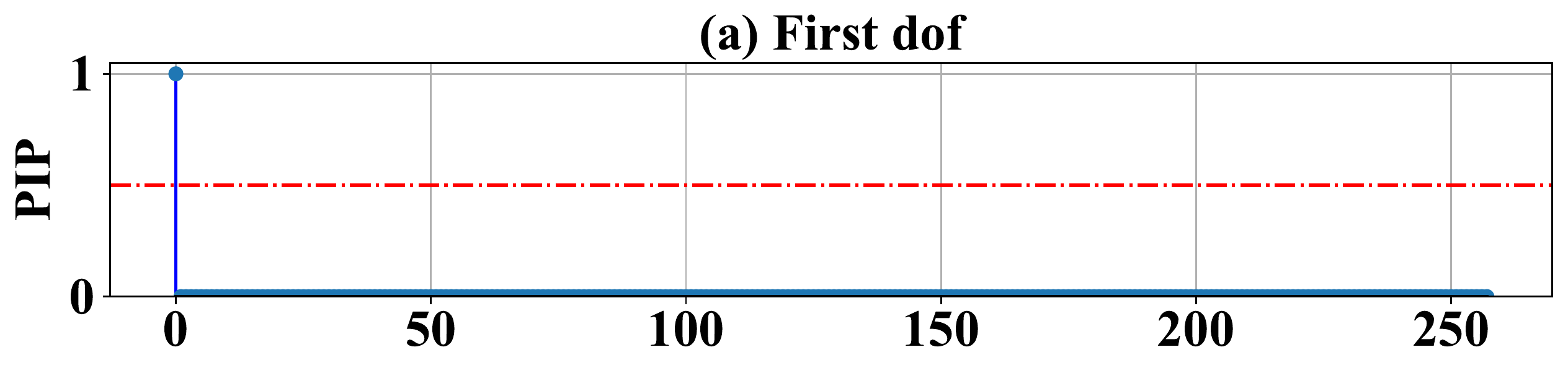}
    \includegraphics[width=0.75\textwidth]{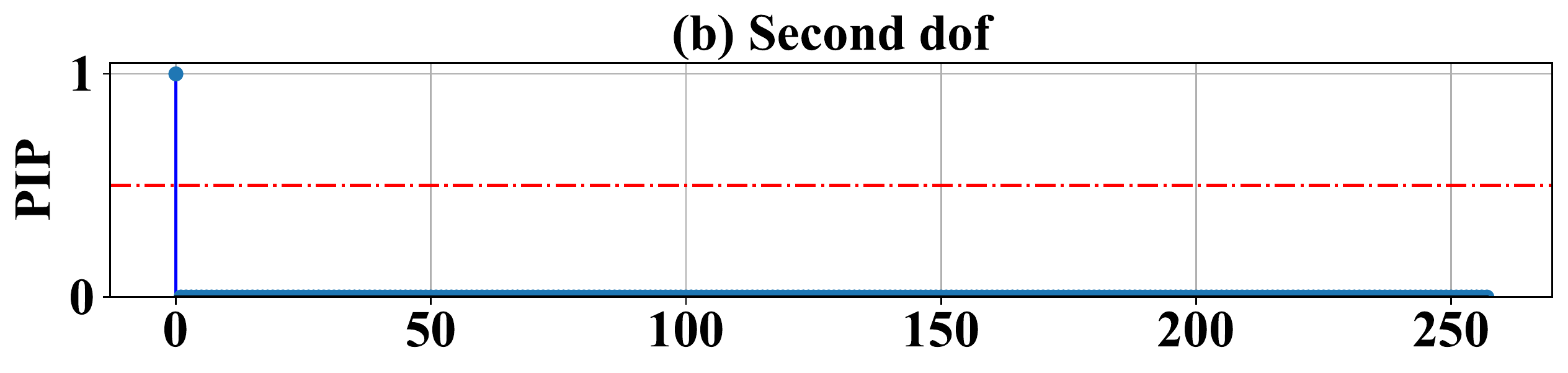}
    \includegraphics[width=0.75\textwidth]{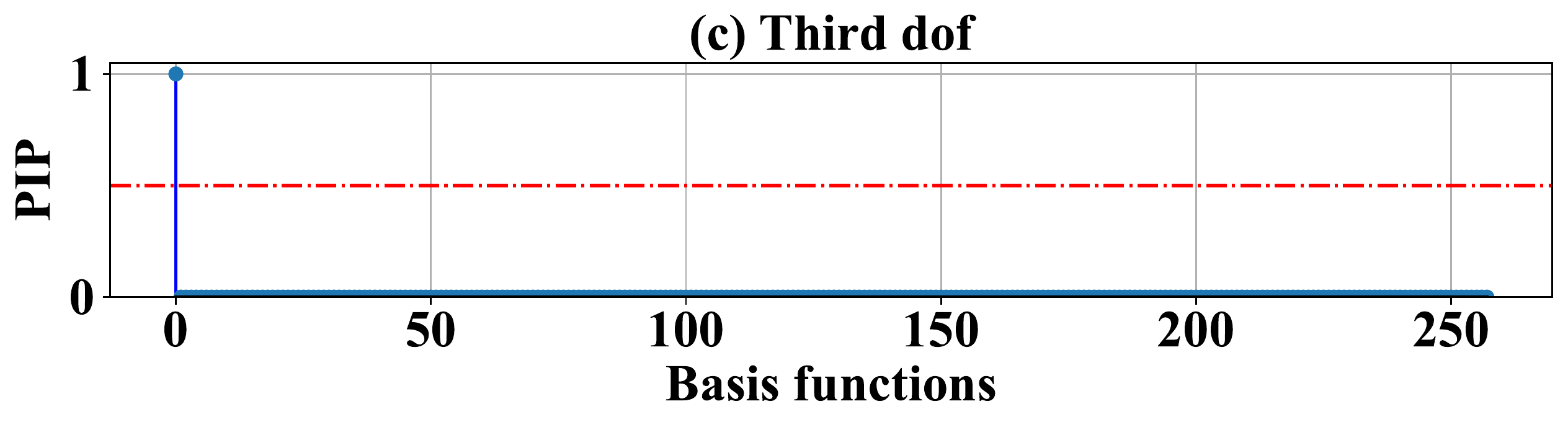}
    \caption{\textbf{Basis function selection for diffusion component of a 3-DOF Duffing oscillator system.} For diffusion terms of 3-DOF non-linear oscillator, the dictionary $\textbf{L} \in \mathbb{R}^{N\times258}$ of 258 basis functions is used, which contains polynomial terms upto fourth order. The basis functions representing the equation are picked by the model with PIP almost equal to 1. For all 3 equations in the diffusion term, the predicted basis function is a constant term. The parameter values associated with the selected basis functions predicted by the model are shown in Table \ref{tab:tab2}}
    \label{fig:3dof_diff}
\end{figure}
\begin{figure}[!ht]
    \centering
    \includegraphics[width=0.75\textwidth]{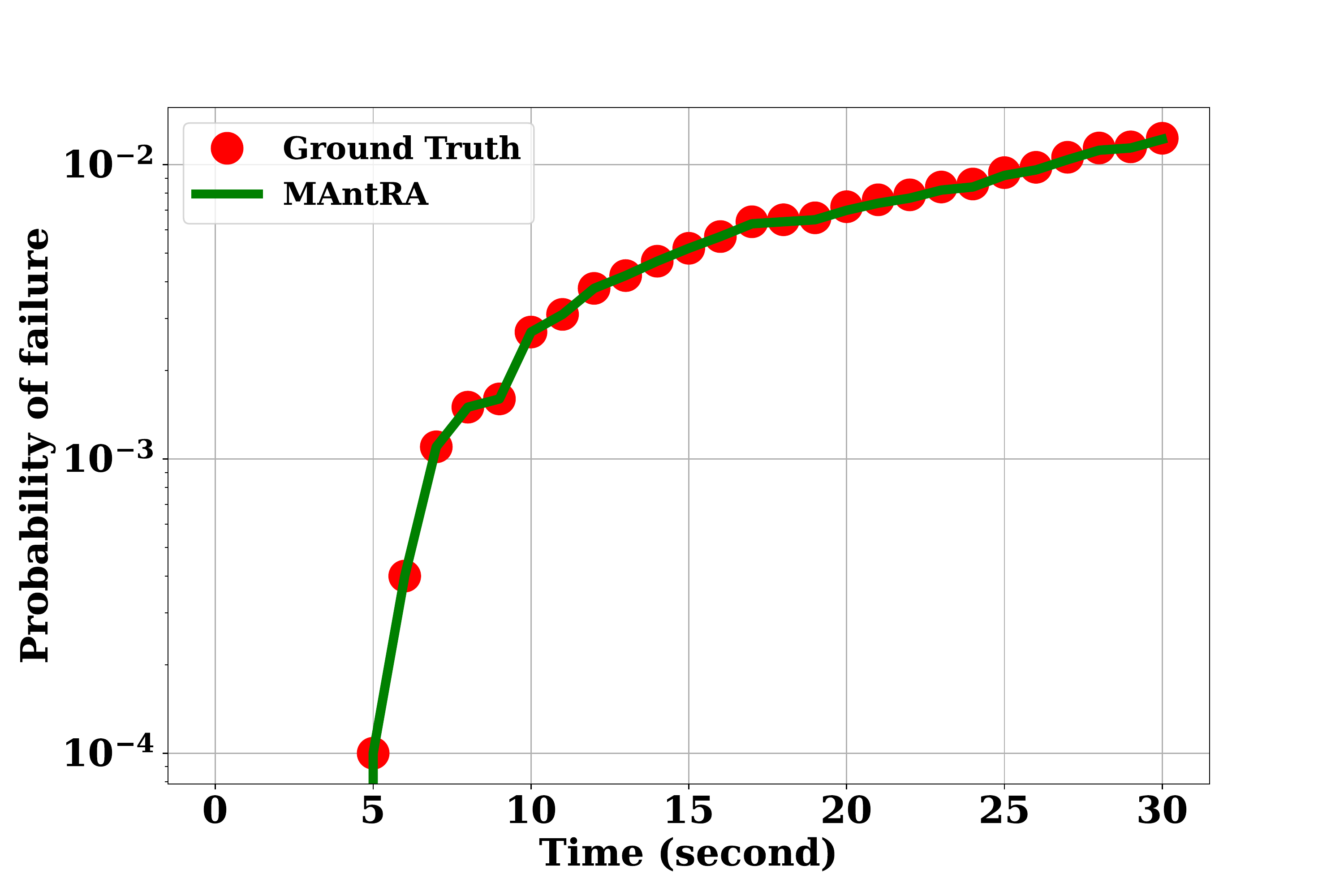}
    \caption{Probability of failure of a 3-DOF Duffing oscillator system with true and discovered models}
    \label{fig:rel_3dof}
\end{figure}
For the $3$-DOF non-linear oscillator problem, we have generated synthetic data using E-M scheme. A time-step of {$\Delta t = 0.001$} is considered and, data corresponding to $t\in[0,1]$s is generated. To track the aleatoric uncertainty, {$N_t = 500$} such realizations were generated. The generated realizations were corrupted with {$5\%$} noise. The proposed MAntRA was employed on the generated data. For this problem ,the dictionary $\mathbf{L} \in \mathbb{R}^{N\times258}$ contains polynomial functions upto order 4, trigonometric functions, signum functions, etc. It is evident from the Fig. \ref{fig:3dof_driftl} and Fig. \ref{fig:3dof_diff} that, all the relevant basis functions for drift and diffusion terms respectively, were identified accurately. The parameters of the identified equations are given in Table \ref{tab:tab2} (last column). Time taken to make the equation discovery here is considerably lesser as compared to the MCMC-based methods such as Gibbs sampling. This clearly indicates that the developed variational Bayesian SDE discovery framework is capable of identifying the governing SDE for different DOFs from  one second of noisy displacement measurements, sampled at a sampling frequency of $1000$Hz.

\begin{table}[!ht]
    \centering
    \caption{System parameters (actual and predicted by the model) for Example problem 2 - 3-DOF Duffing oscillator}
    \label{tab:tab2}
    \begin{threeparttable}
        \begin{tabular}{c|c|c}
        \hline
        Parameter & Actual value & Predicted value\\
        \hline
        Damping (Ns/m) & $\hat{c}_i=2$, $i=1,2,3$ & $\hat{c}_1 =2.001$, $\hat{c}_2=1.999$, $\hat{c}_3=2.011$ \\
        Stiffness (N/m) & $\hat{k}_1=1000$, $\hat{k}_2=2000$, $\hat{k}_3=3000$ & $\hat{k}_1 =1000.00$, $\hat{k}_2=1999.67$, $\hat{k}_3=2999.87$ \\
        Diffusion  & $\hat{\sigma}_i=1$, $i=1,2,3$ & $\hat{\sigma}_1 =0.99$, $\hat{\sigma}_2=1.06$, $\hat{\sigma}_3=0.94$\\
        Nonlinear & $\hat{\alpha}_i=100000$, $i=1,2,3$ & $\hat{\alpha}_1 =100408.8$, $\hat{\alpha}_2=100234.4$\\
        & & $\hat{\alpha}_3=100349.8$\\
        \hline
        \end{tabular}
    \end{threeparttable}
    \begin{tablenotes}
        \item Note: Here $\hat{[(\cdot)]}$ denotes the mass normalized parameters of the system.
    \end{tablenotes}
\end{table}
After validating the equation discovery using the proposed MAntRA, we move onto solving the time dependent reliability analysis problem for the considered system. As discussed in the section \ref{sec:system1} the limit state function $\mathcal J \left(\zeta \right)$ as given in Eq. \eqref{eq:limit-state1} is used to compute the probability of failure of the system using Eq. \eqref{eq:pf}. The threshold $X_t$ for this problem is taken as $0.18$. The results of probability of failure as shown in the plot in Fig. \ref{fig:rel_3dof} indicate that, the probability of failure using the proposed MAntRA match exactly with the reference solution obtained by using True SDE. MAntRA yields excellent results for $t=30$s, which is a distant location from the training window.

\subsection{5-DOF linear structural system with tuned mass damper}\label{sec:system3}

\begin{figure}[!ht]
    \centering
    \includegraphics[width=0.7\textwidth]{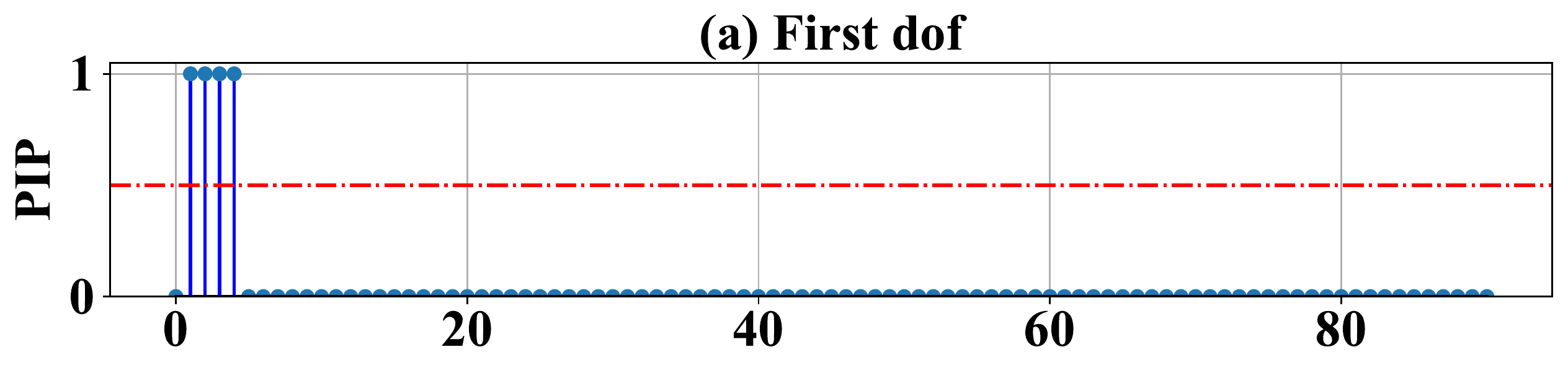}
    \includegraphics[width=0.7\textwidth]{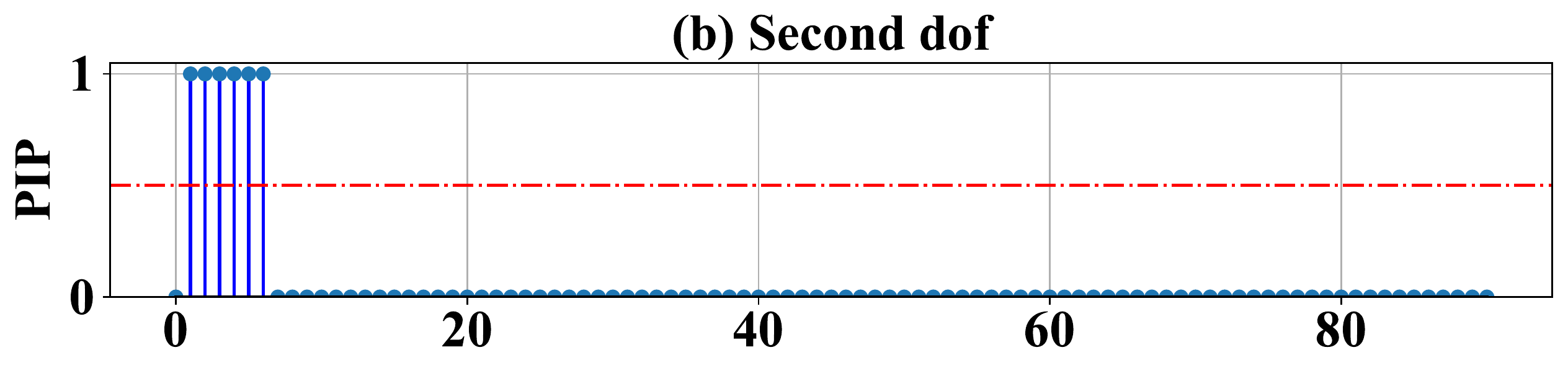}
    \includegraphics[width=0.7\textwidth]{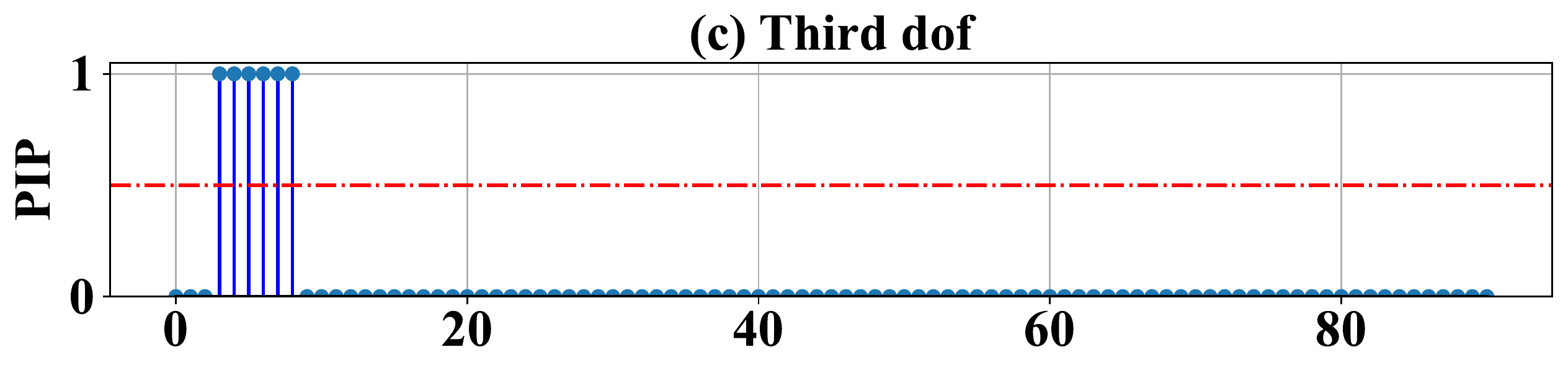}
    \includegraphics[width=0.7\textwidth]{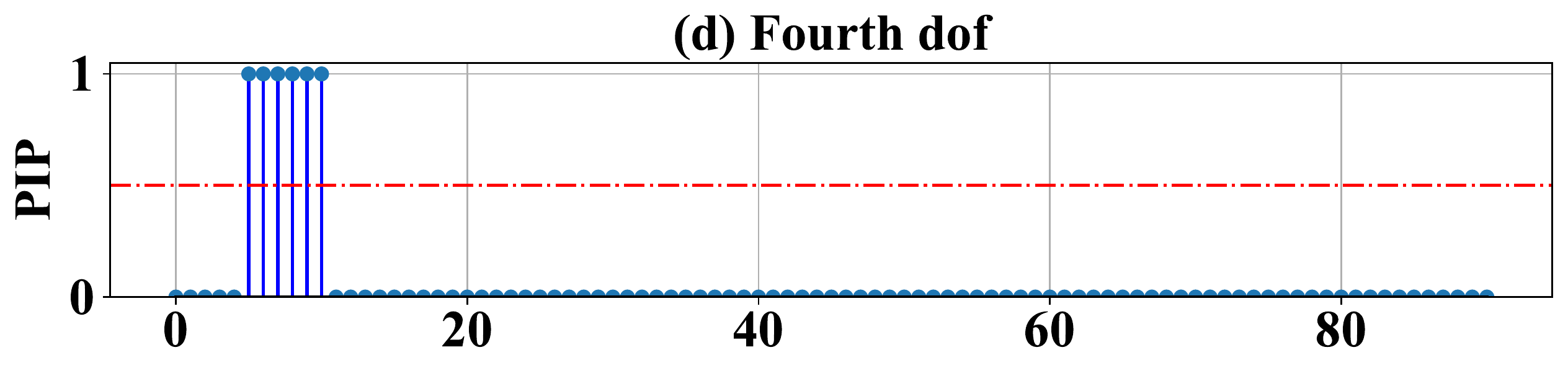}
    \includegraphics[width=0.7\textwidth]{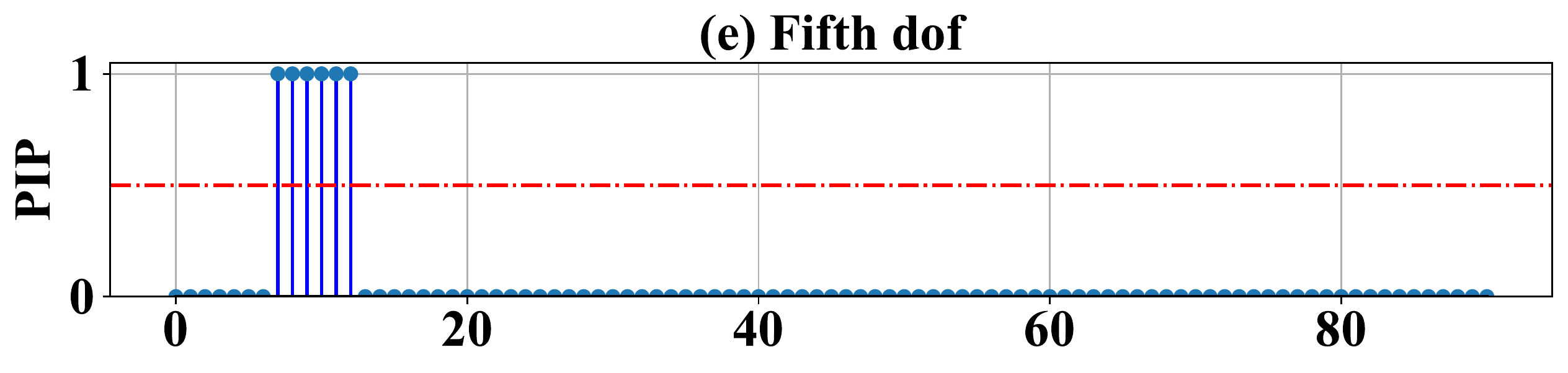}
    \includegraphics[width=0.7\textwidth]{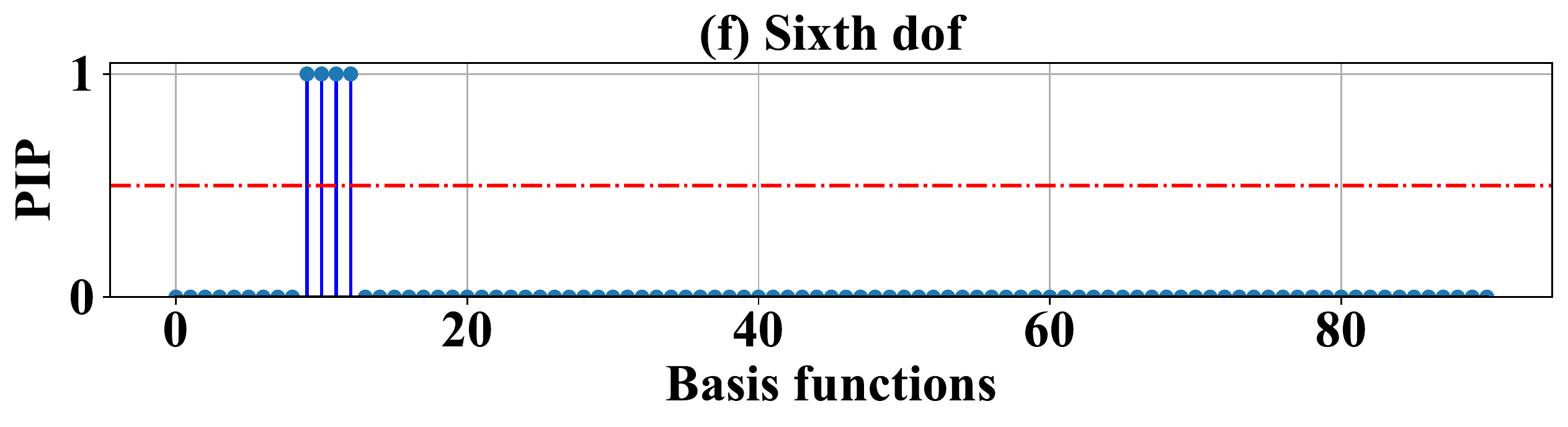}
    \caption{\textbf{Identification of TMD drift terms.} For TMD problem the dictionary $\textbf{L} \in \mathbb{R}^{N\times92}$ of 92 basis functions is used. There are 5 degrees of freedom, and the 6$^{\text{th}}$ is the TMD, all 6 equations of the system are identified correctly with \textbf{PIP} almost equals to 1 as shown in Fig. \ref{fig:tmd_drift}. The parameter values associated with the selected basis functions predicted by the model are shown in Table \ref{tab:tab3}.}
    \label{fig:tmd_drift}
\end{figure}
 
\begin{figure}[ht!]
    \centering
    \includegraphics[width=0.7\textwidth]{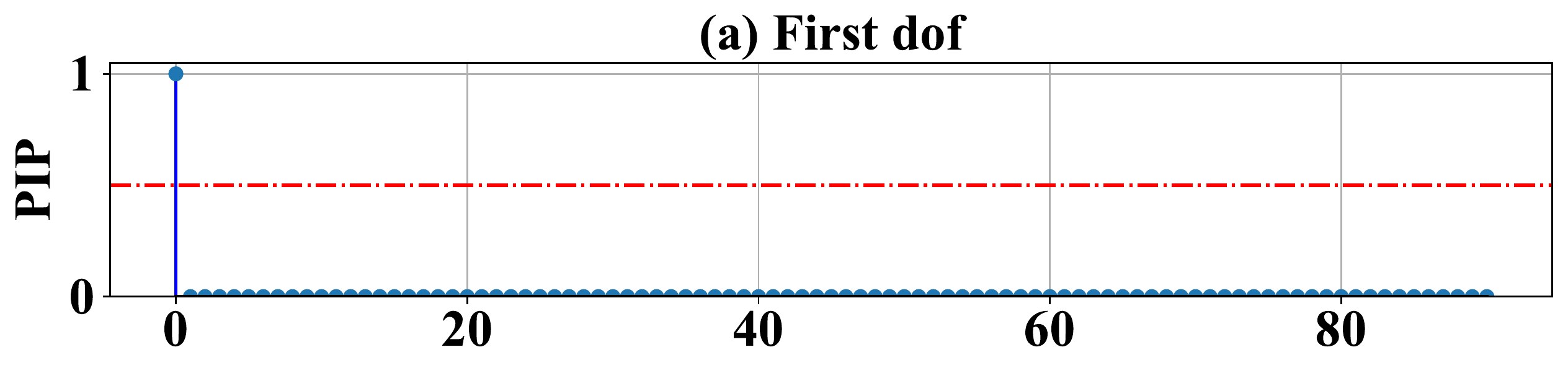}
    \includegraphics[width=0.7\textwidth]{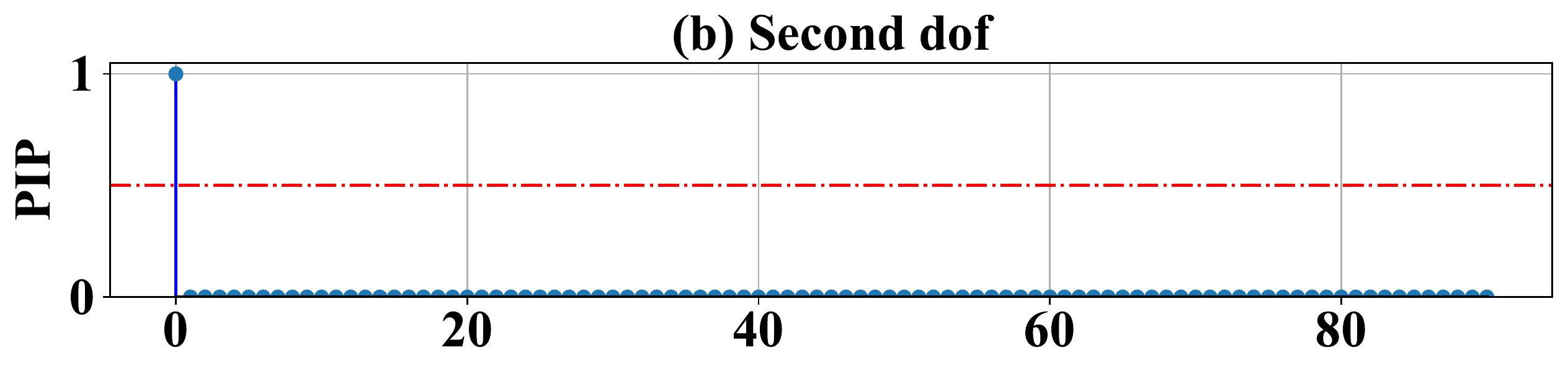}
    \includegraphics[width=0.7\textwidth]{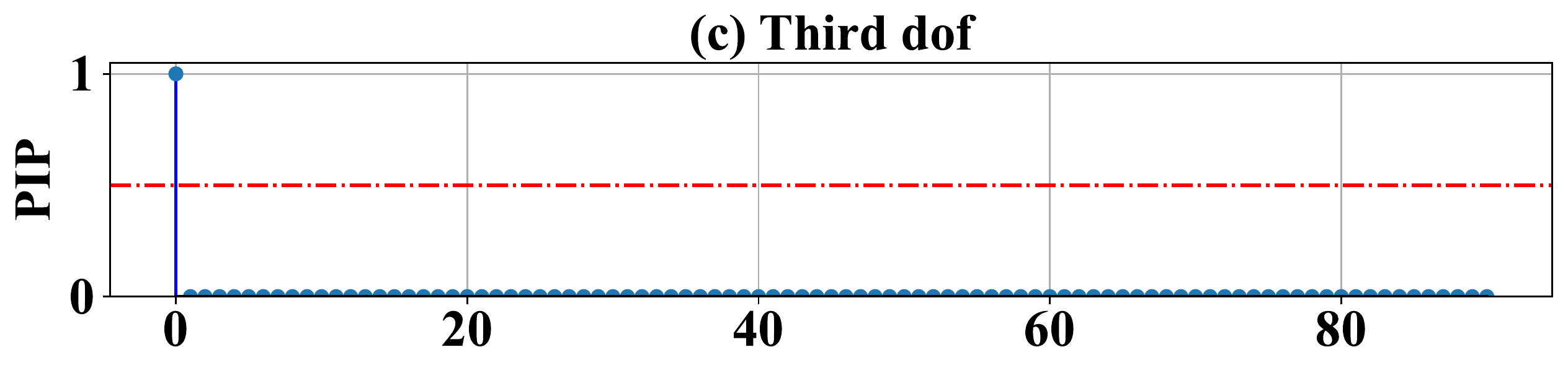}
    \includegraphics[width=0.7\textwidth]{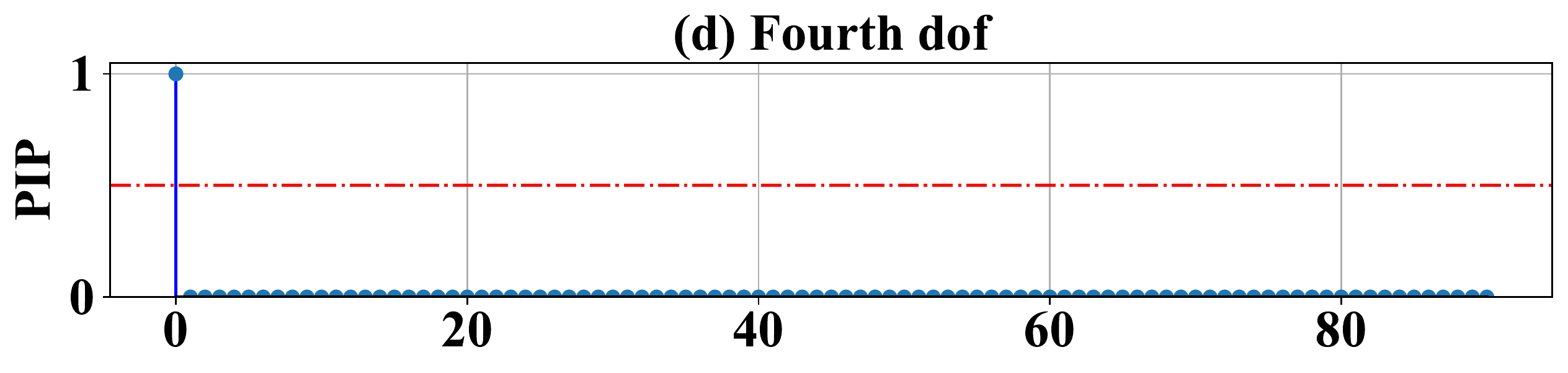}
    \includegraphics[width=0.7\textwidth]{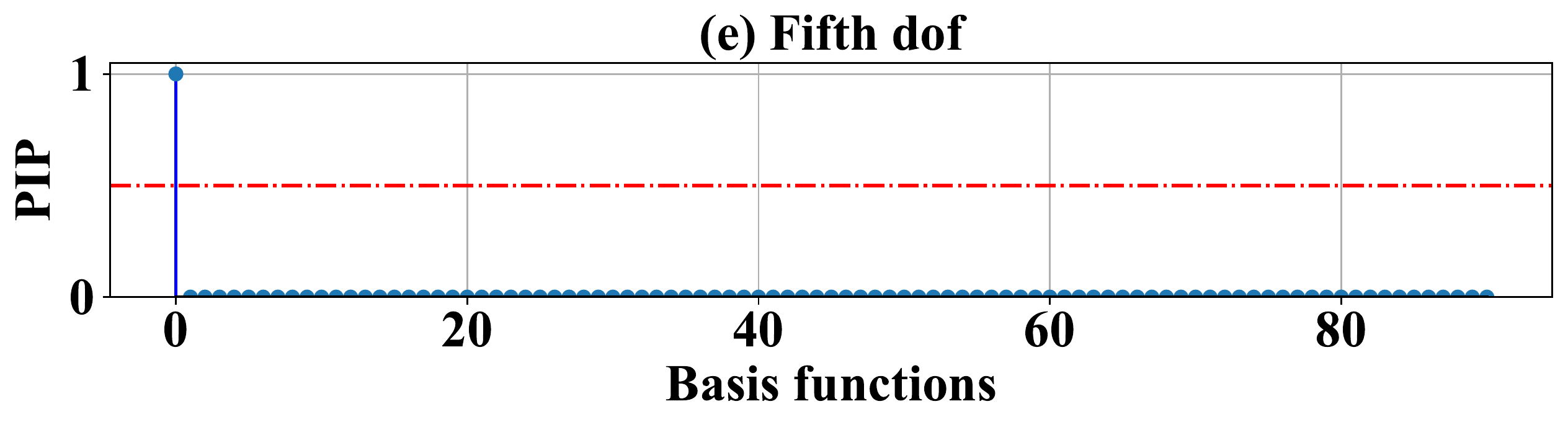}
    \caption{\textbf{Identification of TMD diffusion terms.} For diffusion terms the dictionary $\textbf{L} \in \mathbb{R}^{N\times92}$ of 92 basis functions is used. There are 5 degrees of freedom for which diffusion term will appear in the equation, all 5 equations of the system are identified correctly with \textbf{PIP} almost equal to 1 as shown in  Fig. \ref{fig:tmd_diff}. The parameter values associated with the selected basis functions predicted by the model are shown in Table \ref{tab:tab3}.}
    \label{fig:tmd_diff}
\end{figure}
\begin{figure}[ht!]
    \centering
    \includegraphics[width=0.75\textwidth]{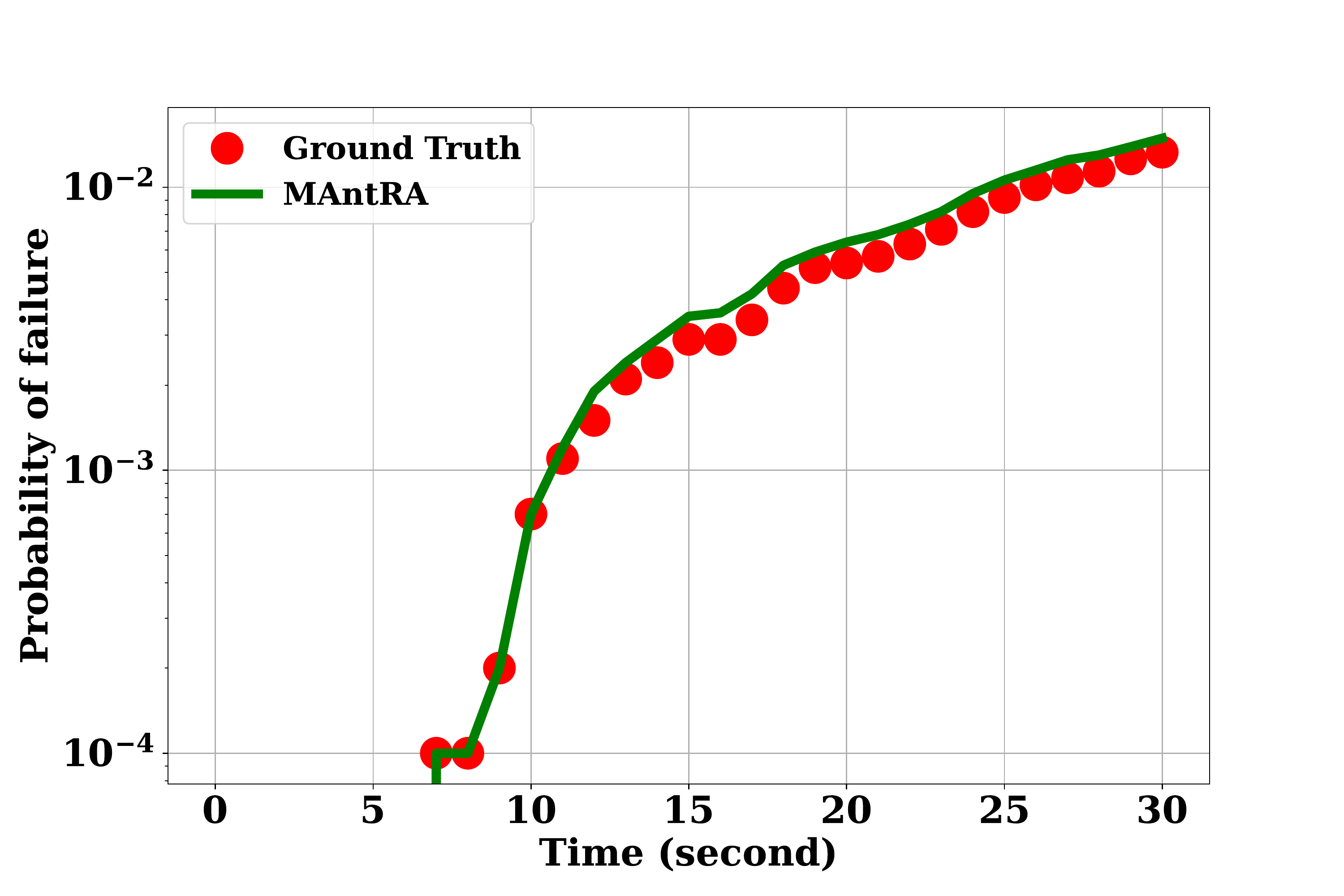}
    \caption{Probability of failure of a linear tuned mass damper system with true and discovered models}
    \label{fig:rel_tmd}
\end{figure}

The third system considered here is a 5-DOF structural system with a tuned mass damper (TMD) devised at the top floor of the structure. TMDs are one of the oldest structural vibration control devices that are in existence \cite{chang1995control,rana1998parametric}. There are a significant number of tall structures that are equipped with different kinds of TMD systems for vibration mitigation under wind and moderate earthquakes. In this work, we are using a 5-DOF dynamical system that has a tuned mass damper at the top floor. The equation of the system can be represented in the form of the spring-mass-dashpot model as follows:

\begin{equation}
\textbf{M}{\ddot{\bm X}}(t)+\textbf{C}{\dot{\bm X}}(t)+ \textbf{K}\bm{X}(t) = \bm{\Sigma}{\dot{\bm B}}(t) \label{eq:TMD}
\end{equation}
where \textbf{M}, \textbf{C}, and \textbf{K} are the $\mathbb{R}^{6 \times 6}$ mass, damping and stiffness matrices, respectively, $\bm{\Sigma} \in \mathbb{R}^{6 \times 6}$ is the diffusion matrix, and $\bm{\dot{B}}{(t)} \in \mathbb{R}^{6}$ is the vector of derivatives of Brownian motion. The values of parameters in the equation are given in Table \ref{tab:tab3}. An appropriate statespace is used to construct the first order It\^{o} SDEs for the differential system.\\
Similar to the other two problems, data is generated using E-M scheme taking the time step as $\Delta{t} = 0.001$. For this problem, the dictionary $\textbf{L} \in \mathbb{R}^{N\times92}$ of 92 basis functions is used, which contains polynomial terms upto order 3. The basis functions in the drift and diffusion term representing the equations are picked by the proposed MAntRA with PIP almost equal to 1, as shown in Fig. \ref{fig:tmd_drift} and Fig. \ref{fig:tmd_diff} respectively. The values of the parameters associated with the basis functions selected in the drift as well as diffusion term are given in Table \ref{tab:tab3} (last column). The estimated parameters match almost exactly with the actual values indicating the capability of developed variational Bayesian SDE discovery framework for identifying the governing SDE from  one second of noisy displacement measurements, sampled at a sampling frequency of $1000$Hz.

\begin{table}[ht!]
    \centering
    \caption{System parameters (actual and predicted by the model) for Example problem 3 - tuned mass damper}
    \label{tab:tab3}
    \begin{threeparttable}
        \begin{tabular}{c|c|c}
        \hline
        Parameter & Actual value & Predicted value\\
        \hline
        Damping (Ns/m) & $\hat{c}_i=2$; $i=1,\ldots,6$ & $\hat{c}_1 =2.05$, $\hat{c}_2=2.02$,$c_3=1.92$ \\
        & & $\hat{c}_4=2.00$, $\hat{c}_5=2.01$, $\hat{c}_6=2.00$\\
        \hline
        Stiffness (N/m) & $\hat{k}_1=1000$,$\hat{k}_2=1500$,$\hat{k}_3=2000$ & $\hat{k}_1 =990.90$, $\hat{k}_2=1513.22$,$\hat{k}_3=1988.77$ \\
        & $\hat{k}_4=2500$,
        $\hat{k}_5=3000$,$\hat{k}_6=300$ & $\hat{k}_4=2504.51$, $\hat{k}_5=2995.49$, $\hat{k}_6=873.39$\\ 
        \hline
        Diffusion & $\hat{\sigma}_i=1$;$i=1,\ldots,5$, $\hat{\sigma}_6=0$ & $\hat{\sigma}_1 =1.00$, $\hat{\sigma}_2=1.01$,$\hat{\sigma}_3=0.97$ \\
        & & $\hat{\sigma}_4=0.99$, $\hat{\sigma}_5=1.01$\\ 
        \hline
        \end{tabular}
    \end{threeparttable}
    \begin{tablenotes}
        \item Note: Here $\hat{[(\cdot)]}$ denotes the mass normalized parameters of the system.
    \end{tablenotes}
\end{table}

After validating the equation discovery using the proposed MAntRA, we move onto solving the time dependent reliability analysis problem for the considered system. As discussed in the section \ref{sec:system1} the limit state function $\mathcal J \left(\zeta \right)$ as given in Eq. \eqref{eq:limit-state1} is used to compute the probability of failure of the system using Eq. \eqref{eq:pf}. The threshold $X_t$ for this problem is taken as $0.0614$. The results of probability of failure as shown in the plot in Fig. \ref{fig:rel_tmd} indicate that, the probability of failure using the proposed MAntRA match exactly with the reference solution obtained by using True SDE. MAntRA yields excellent results for $t=30$s, which is a distant location from the training window.

\section{Discussion and Conclusion}\label{conclusion}
In this paper, we have presented MAntRA, a computationally efficient approach to solve the reliability analysis problem where the system model is taken to be a priori unknown, and we only have access to displacement measurements of the system. Conventional reliability analysis methods cannot be directly applied to cases where the model is not known, as a model is required to evaluate the limit state function. The proposed framework considers stochastic models, that is models where the input is modeled as a Brownian motion and makes use of stochastic calculus, sparse learning, and Bayesian variable selection to determine the structure and the parameters of the drift and diffusion terms. While any data-driven black-box model could have been used to represent these terms, the poor explainability (and as a consequence generalizability) of the black-box models could lead to incorrect reliability values.

To promote interpretability in deducing the model, a library of manually-designed candidate functions is constructed and then relevant functions from the library are selected by enforcing sparsity in the corresponding parameters. The ensuing problem of selecting a model structure and determining its parameters reduces to a sparse linear regression problem. Employing a Bayesian framework, a strong sparsity-promoting spike and slab prior is combined with variational Bayes (VB) algorithm to determine the drift and diffusion terms of the underlying SDE. In case the library of functions does not include the true terms present in the actual model, the algorithm will select a set of correlated basis functions from the library, in which case the model becomes a surrogate model. The VB algorithm is chosen over a more accurate MCMC technique because it is computationally very efficient, and the results obtained were reasonably close to those obtained from MCMC methods. With the discovered drift and diffusion terms, we perform the reliability analysis. For a given time, the probability of failure is computed based on the number of times the
system exceeds the threshold displacement. To check the efficacy of the proposed approach, we have taken three example problems to solve. In the first problem, Duffing oscillator which has cubic non-linearity has been solved using VB, and the probability of failure is computed. In the second problem, the reliability analysis of the non-linear system with three DOF is solved. As a third problem, we have taken a linear tuned mass damper system. it can be seen from the results that the probability of failure of the discovered systems is almost similar to that of the original system. The results generated indicate the utility and potential of the proposed approach.

\section*{Acknowledgements}
T. Tripura acknowledges the financial support received from the Ministry of Education (MoE), India, in the form of the Prime Minister's Research Fellowship (PMRF). S. Chakraborty acknowledges the financial support received from Science and Engineering Research Board (SERB) via grant no. SRG/2021/000467 and seed grant received from IIT Delhi.

\section*{Code availability}
Upon acceptance, all the source codes to reproduce the results in this study will be made available to the public on GitHub by the corresponding author.

\section*{Competing interests} 
The authors declare no competing interests.


\end{document}